\documentclass{emulateapj}
\usepackage{epsfig,natbib}
\usepackage{graphicx}
\newcommand{\ms}{\mbox{m\,s$^{-1}~$}}

\newcommand{\mse}{\mbox{m\,s$^{-1}$}}

\newcommand{\msun}{M$_{\odot}~$}

\newcommand{\rsun}{R$_{\odot}~$}

\newcommand{\mearth}{$M_\earth$~}
\newcommand{\mearthe}{$M_\earth$}

\newcommand{\rearthe}{$R_\earth$}

\newcommand{\mstar}{\ensuremath{M_{\star}}}
\newcommand{\rstar}{\ensuremath{R_{\star}}}

\newcommand{\feh}{\ensuremath{[\mbox{Fe}/\mbox{H}]}}

\newcommand{\rphk}{\ensuremath{R'_{\mbox{\scriptsize HK}}}}
\newcommand{\shk}{\ensuremath{S_{\mbox{\scriptsize HK}}}}
\newcommand{\lrphk}{\ensuremath{\log{\rphk}}}
\newcommand{\caii}{\ion{Ca}{2} H \& K}

\newcommand{\msini}{\ensuremath{M \sin i}}

\newcommand{\teff}{\ensuremath{T_{\rm eff}}}

\newcommand{\logg}{\ensuremath{\log{g}}}

\newcommand{\bjdtdb}{\ensuremath{\rm {BJD_{TDB}}}}

\newcommand{\bmass}{8.7 \mearthe}
\newcommand{\cmass}{7.9 \mearthe}
\newcommand{\dmass}{6.4 \mearthe}

\newcommand{\bper}{5.4 days}
\newcommand{\cper}{15.3 days}
\newcommand{\dper}{24.5 days}
\newcommand{\longper}{$\sim$2400 days}
\newcommand{\actper}{40.8 days}
\newcommand{\aactper}{17.1 days}
\newcommand{\binapf}{80}
\newcommand{\binkeck}{281}
\newcommand{\numapf}{109}
\newcommand{\numkeck}{797}

\shortauthors{Fulton {et~al.}}
\shorttitle{Three Super-Earths}
\submitted{In Preparation for  ApJ}
\begin{document}
\pagenumbering{arabic}

\title{Three Super-Earths Orbiting HD 7924\altaffilmark{1,2}}
\author{
Benjamin J.\ Fulton\altaffilmark{3,7,*},
Lauren M.\ Weiss\altaffilmark{4,8},
Evan Sinukoff\altaffilmark{3,9},
Howard Isaacson\altaffilmark{4},
Andrew W. Howard\altaffilmark{3},
Geoffrey W. Marcy\altaffilmark{4},
Gregory W. Henry\altaffilmark{5},
Bradford P. Holden\altaffilmark{6},
Robert I. Kibrick\altaffilmark{6}
}
\altaffiltext{1}{We refer to this planetary system as the ``Levy Planetary System" to
		     thank Ken and Gloria Levy for their generous contribution to the APF facility.
                     } 
\altaffiltext{2}{Based on observations obtained at the W.\,M.\,Keck Observatory, 
                      which is operated jointly by the University of California and the 
                      California Institute of Technology.  
                      Keck time was granted for this project by the University of Hawai`i, the University of California, and NASA.
                      } 
\altaffiltext{3}{Institute for Astronomy, University of Hawai`i, 2680 Woodlawn Drive, Honolulu, HI 96822, USA} 
\altaffiltext{4}{Department of Astronomy, University of California, Berkeley, CA 94720, USA}
\altaffiltext{5}{Center of Excellence in Information Systems, Tennessee State University, Nashville, TN 37209, USA}
\altaffiltext{6}{University of California Observatories, University of California, Santa Cruz, CA 95064, USA}
\altaffiltext{7}{National Science Foundation Graduate Research Fellow}
\altaffiltext{8}{Ken and Gloria Levy Graduate Student Research Fellow}
\altaffiltext{9}{Natural Sciences and Engineering Research Council of Canada Graduate Student Fellow}
\altaffiltext{*}{bfulton@hawaii.edu}

\begin{abstract}
We report the discovery of two super-Earth mass planets orbiting the nearby K0.5 dwarf HD 7924 which was previously known to host one small planet. The new companions have masses of 7.9 and 6.4 M$_\oplus$, and orbital periods of 15.3 and 24.5 days. We perform a joint analysis of high-precision radial velocity data from Keck/HIRES and the new Automated Planet Finder Telescope (APF) to robustly detect three total planets in the system. We refine the ephemeris of the previously known planet using five years of new Keck data and high-cadence observations over the last 1.3 years with the APF. With this new ephemeris, we show that a previous transit search for the inner-most planet would have covered 70\% of the predicted ingress or egress times. Photometric data collected over the last eight years using the Automated Photometric Telescope shows no evidence for transits of any of the planets, which would be detectable if the planets transit and their compositions are hydrogen-dominated. We detect a long-period signal that we interpret as the stellar magnetic activity cycle since it is strongly correlated with the Ca II H \& K activity index. We also detect two additional short-period signals that we attribute to rotationally-modulated starspots and a one month alias. The high-cadence APF data help to distinguish between the true orbital periods and aliases caused by the window function of the Keck data. The planets orbiting HD 7924 are a local example of the compact, multi-planet systems that the Kepler Mission found in great abundance.
\end{abstract}


\section{Introduction}
\label{sec:intro}

The archetypal planets of our Solar System---Jupiter the gas giant, Neptune the ``ice'' giant, and Earth the terrestrial planet---represent 
an incomplete inventory of the planet types in our galaxy.  
We are locally impoverished in ``super-Earths,'' the broad category of planets intermediate in size and mass between Earth and Neptune.
Doppler searches of nearby stars showed that super-Earth planets in close-in orbits are plentiful \citep{Howard10,Mayor11}.  
Results from the Kepler mission confirmed and refined our knowledge of the size and orbital period distribution of these  and other planets  
\citep{Howard12,Petigura13,Fressin13}.
These planets have a wide range of bulk densities \citep{Marcy14}, suggesting a diversity of compositions spanning  
rocky planets with negligible atmospheres \citep{Howard13,Pepe13} 
to puffy planets with thick gas envelopes \citep{Kipping14}.
Intermediate planets with densities of $\sim$\,3 g\,cm$^{-3}$ are consistent with a broad range of interior structures and atmosphere sizes.    
Planets smaller than $\sim$1.6 Earth radii (\rearthe) are more likely to have a high density 
and presumed rocky composition \citep{Weiss14,Rogers14}.

The large population of super-Earths orbiting close to their host stars was a surprise.  
Population synthesis models of planet formation had predicted that such systems would be rare \citep{Ida04,Mordasini09}.  
Planet cores were expected to mostly form  beyond the ice line and rarely 
migrate to close orbits unless they first grew to become gas giants.  
Nevertheless, close-in, low-mass planets are common and often appear in compact multi-planet systems \citep{Lissauer11,Fang12}.  
Theoretical models are catching up, with refinements to the disk migration and multi-planet dynamics in the population 
synthesis family of models \citep{Ida10,Alibert13, Schlichting14, Lee14}.  
A new class of ``in situ'' formation models have also been proposed in which systems of super-Earths and Neptunes 
emerge naturally from massive disks \citep{Hansen12,Chiang13}.

The Eta-Earth Survey \citep{Howard10} at Keck Observatory played a important role in the 
discovery that super-Earths are abundant.  Using the HIRES spectrometer, 
our team searched for planets in a volume-limited sample of 166 nearby G and K dwarf stars.  
Our search yielded new planets \citep{Howard09,Howard11a,Howard11b,Howard14} and 
detection limits for each star.  
Putting these together, we measured the prevalence of planets in close-in orbits as a function of planet mass (\msini).  
This mass function rises steeply with decreasing mass: planets in the mass range 3--10 \mearthe\ 
are about twice as common as 10--30 \mearthe\ planets.

The first low-mass planet discovered in the Eta-Earth Survey was HD 7924b \citep[][H09 hereafter]{Howard09}, a super-Earth with a 
mass of \bmass\ and an orbital period of \bper.  
We have continued to observe HD 7924 and other Eta-Earth Survey stars using HIRES.  
These additional measurements probe smaller masses and larger star-planet separations.  
We have also started observing a subset of the Eta-Earth Survey stars with the Automated Planet Finder 
\citep[APF;][]{Vogt14}, a new telescope at Lick Observatory.  
APF is a robotic 2.4-m telescope designed exclusively for Doppler discovery of exoplanets.  
It feeds the high-resolution Levy Spectrometer \citep{Radovan10} that uses 
an iodine reference spectrum to calibrate the wavelength scale and point spread function \citep{Butler96b}, 
achieving a Doppler precision similar to HIRES while running without human intervention during an observing night.  
APF exploits high measurement cadence (nearly nightly) to 
disentangle the complicated low-amplitude signals of multi-planet systems in the face of stellar activity. 

In this paper we announce two additional super-Earths orbiting HD 7924 based on RVs from the Keck-HIRES and APF-Levy spectrometers.  
We describe the properties of the  star HD 7924 in Sec.\ \ref{sec:stellar_props} 
and our Doppler measurements from APF/Levy and Keck/HIRES in Sec.\ \ref{sec:measurements}.   
Our analysis of the RV data, including discovery of the three Keplerian signals and consideration of 
false alarm probabilities, alias periods, and chromospheric activity,
is described in Sec.\ \ref{sec:keplerian}.  
We conclude with a discussion and summary in Sec.\ \ref{sec:discussion}.

\section{Stellar Properties}
\label{sec:stellar_props}

HD 7924, also known as HIP 6379 or GJ 56.5, is a nearby \citep[16.82 pc;][]{vanLeeuwen07} and bright K0.5V dwarf star \citep{vonBraun14}. It is slightly metal poor relative to the Sun and hosts one previously known planet. It is relatively inactive with $\log{\rphk}=-4.89$ \citep{Isaacson10}, but we do detect some low-level chromospheric activity nonetheless (see Section \ref{sec:activity}). We list our adopted stellar parameters in Table \ref{tab:stellar_params}.

Most of our spectroscopically-derived stellar parameters are adopted from H09, which were originally derived by \citet{Valenti05} using the Spectroscopy Made Easy (SME) LTE spectral
synthesis code. However, HD 7924 has been the focus of several studies since the discovery of HD 7924b. \citet{Santos13} performed a uniform analysis of 48 planet-hosting stars. They find $\teff=5133\pm68$ K, $\logg=4.46\pm0.12$, and $\feh=-0.22\pm0.04$ for this star.
\citet{vonBraun14} use the empirical relations of \citet{Boyajian12} to calculate a mass of 0.81 \msun for HD 7924 with an error estimate of 30\%. We computed the stellar mass and radius from \teff, \logg, and \feh\ using the \citet{Torres10} relations and found $0.81 \pm 0.02$ \msun and $0.75 \pm 0.03$ \rsun. All of these values are within 1-$\sigma$ of our adopted values.

HD 7924 was observed by \citet{vonBraun14} using long-baseline interferometry on the Georgia State University Center for High Angular Resolution Astronomy (CHARA) Array \citep{tenBrummelaar05}. We adopt their value for \teff\ of $5075\pm83$ K that they obtained by fitting the spectral energy distribution from optical through infrared wavelengths. They also obtained precise values for the luminosity and radius that we list in Table \ref{tab:stellar_params}. These values are consistent within $1-\sigma$ with the values adopted in H09 that were based on fitting stellar evolution models with spectroscopic parameters. \citet{Mason11} conducted high-resolution speckle imaging of HD 7924 and detected no companions within three V magnitudes of HD 7924 with separations between 0\farcs03 and 1\farcs5.

\begin{deluxetable}{lcc}
\tabletypesize{\footnotesize}
\tablecaption{Adopted Stellar Properties of HD 7924}
\tablewidth{0pt}
\tablehead{
  \colhead{Parameter}   & 
  \colhead{HD 7924} &
  \colhead{Source} 
}
\startdata
Spectral type ~~~~~~& K0.5V  & \citet{vonBraun14} \\
$B-V$ (mag) & 0.826  &  H09  \\
$V$ (mag)   & 7.185  & H09 \\
$J$  (mag)  & 5.618 $\pm$ 0.026  & \citet{Cutri03}  \\
$H$  (mag)  & 5.231 $\pm$ 0.033  & \citet{Cutri03} \\
$K$ (mag)   & 5.159 $\pm$ 0.020 & \citet{Cutri03}  \\
Distance (pc) & 16.82 $\pm$ 0.13  & \citet{vanLeeuwen07}  \\
$T_\mathrm{eff}$ (K) &  5075 $\pm$ 83  & \citet{vonBraun14}  \\
log\,$g$ (cgs) & 4.59$^{+0.02}_{-0.03}$  &  H09 \\
\feh\ (dex) &  $-$0.15 $\pm$ 0.03 &  H09  \\
$v$\,sin\,$i$ (km\,s$^{-1}$) &  $1.35\pm0.5$  & H09 \\
$L_{\star}$ ($L_{\sun}$) & 0.3648 $\pm$ 0.0077 & \citet{vonBraun14}  \\
\mstar\ ($M_{\sun}$) & $0.832^{+0.022}_{-0.036}$ & \citet{Takeda07} \\
\rstar\ ($R_{\sun}$) & 0.7821 $\pm$ 0.0258 &  \citet{vonBraun14}   \\
\lrphk  &  $-$4.89  & \citet{Isaacson10} \\
$S_\mathrm{HK}$  & 0.20 & \citet{Isaacson10} \\
\enddata
\label{tab:stellar_params}
\end{deluxetable}
\vspace{20pt}

\section{Measurements}
\label{sec:measurements}

\subsection{Keck/HIRES Spectroscopy}
\label{sec:keckdata}
We collected 599 new RV measurements for HD 7924 over the last 5 years since the discovery of HD 7924b (H09) using the HIRES spectrograph on the Keck I telescope \citep{Vogt94}. When this new data is combined with the data from H09 and APF we have over 10 years of observational baseline (see Figure \ref{fig:fitplot}a). Our data collection and reduction techniques are described in detail in H09. On Keck/HIRES we observe the star through a cell of gaseous iodine in order to simultaneously forward model the instrumental line broadening function (PSF) and the subtle shifts of the stellar lines relative to the forest of iodine lines. The HIRES detector was upgraded in August of 2004 and the RV zero-point between the pre-and post-upgrade data may not necessarily be the same. For this reason, we allow for separate RV zero-points for the pre-upgrade, post-upgrade, and APF data sets. All RV measurements and associated \caii\ \shk\ activity indices (the ratio of the flux in the cores of the H \& K lines to neighboring continuum levels) are listed in Table \ref{tab:rv}.

\subsection{APF/Levy Spectroscopy}
The Automated Planet Finder is a 2.4 m f/15 Cassegrain telescope built by Electro-Optical Systems Technologies housed in an IceStorm-2 dome located at Lick Observatory atop Mount Hamilton, 20 miles east of San Jose, California. The telescope operates completely unattended using a collection of Python, Tcl, bash, and csh scripts that interact with the lower level software operating on the Keck Task Library keyword system \citep{Lupton93}. Every scheduled observing night, our automation software queries an online Google spreadsheet (that serves as our target database) and creates two observing plans: one for good conditions and the other for poor. Weather permitting, shortly after sunset the observatory opens automatically and determines whether to use the observing plan designed for good or poor conditions by monitoring a bright star. Once the observing plan is started, the high-level software continues to monitor the conditions and adjusts the schedule or switches observing plans if necessary. Observations of our targets (primarily for radial velocity measurements) continue throughout the night until conditions deteriorate too much to remain open or the until morning 9-degree twilight.

The Levy Spectrograph is a high-resolution slit-fed optical echelle spectrograph mounted at one of the two Nasmyth foci of the APF designed specifically for the detection and characterization of exoplanets \citep{Vogt14a, Radovan14, Burt14}. Each spectrum covers a continuous wavelength range from 3740 \AA\ to 9700 \AA. We observed HD 7924 using a 1\farcs0 wide decker for an approximate spectral resolution of $R=100,000$. Starlight passes through a cell of gaseous iodine that serves as a simultaneous calibration source for the instrumental point spread function (PSF) and wavelength reference. In addition, we collected a high signal-to-noise spectrum through the 0\farcs5 wide decker ($R=150,000$) with the iodine cell out of the light path. This spectrum serves as a template from which we measure the relative doppler shifts of the stellar absorption lines with respect to the iodine lines while simultaneously modeling the PSF and wavelength scale of each spectrum. APF guides on an image of the star before the slit using a slanted, uncoated glass plate that deflects 4\% of the light to a guide camera. Photon-weighted times of mid-exposure are recorded using a software-based exposure meter based on guide camera images that monitors the sky-subtracted light entering the slit during an exposure \citep{Kibrick06}.

We measure relative radial velocities (RVs) using a Doppler pipeline descended from the iodine technique in Butler et al. (1996).  For the APF, we forward-model 848 segments of each spectrum between 5000-6200 \AA.  The model consists of a stellar template spectrum, an ultra high-resolution Fourier transform spectrum of the iodine absorption of the Levy cell, a spatially-variable PSF, a wavelength solution, and RV.  We estimate RV uncertainties (Table \ref{tab:rv}) as the uncertainty on the mean RV from the large number of spectral segments.  Well-established RV standard stars have a RV scatter of $\sim$2--3 \ms based on APF measurements.  This long-term scatter represents a combination of photo-limited uncertainties, stellar jitter, and instrumental systematics. We collected a total of \numapf\ RV measurements of HD 7924 on \binapf\ separate nights (post outlier rejection) over a baseline of 1.3 years with a typical per-measurement uncertainty of 2.0 m s$^{-1}$.

For both the APF and Keck data, Julian dates of the photon-weighted exposure mid-times were recored during the observations, then later converted to Barycentric Julian date in the dynamical time system (\bjdtdb) using the tools of \citet{Eastman10}\footnote{IDL tools for time systems conversion; http://astroutils.astronomy.ohio-state.edu/time/.}.

We rejected a handful of low signal-to-noise ratio spectra (S/N $< 70$ per pixel) and measurements with uncertainties greater than nine times the median absolute deviation of all measurement uncertainties relative to the median uncertainty for each instrument. This removed a total of 11 RVs out of the 906 total measurements in the combined data set (Keck plus APF).

After outlier rejection we bin together any velocities taken less than 0.5 days apart on a single telescope. Since data taken in short succession are likely affected by the same systematic errors (e.g. spectrograph defocus), these measurements are not truly independent. Binning helps to reduce the effects of time-correlated noise by preventing multiple measurements plagued by the same systematic errors from being given too much weight in the Keplerian analysis. When the data are binned together and the uncertainty for the resulting data point is added in quadrature with the stellar jitter (an additional error term that accounts for both stellar and instrumental systematic noise), the binned data point receives only as much weight as a single measurement. While this likely reduces our sensitivity slightly we accept this as a tradeoff for more well-behaved errors and smoother $\chi^2$ surfaces. An independent analysis using the unbinned data in Section \ref{sec:alias} finds three planets having the same orbital periods, eccentricities, and masses within 1-$\sigma$ as those discovered by analyzing the binned data.

\begin{deluxetable}{ccccc}
\tabletypesize{\footnotesize}
\tablecaption{Radial Velocities of HD 7924}
\tablewidth{245pt}
\tablehead{ 
    \colhead{\bjdtdb}               & \colhead{RV}  & \colhead{Uncertainty} & \colhead{Instrument\tablenotemark{1}} &  \colhead{\ensuremath{S_{\mbox{\scriptsize HK}}}} \\
    \colhead{(-- 2440000)}  & \colhead{(\mse)}            & \colhead{(\mse)}       & \colhead{} & \colhead{}
}
\startdata
12307.77162 & 5.23 & 1.31 & k & \nodata  \\
12535.95639 & 3.24 & 1.16 & k & \nodata  \\
13239.08220 & -3.00 & 0.91 & j & 0.218   \\
13338.79766 & 3.83 & 1.08 & j & 0.227 \\
16505.99731 & -7.97 & 1.46 & a & \nodata \\
16515.90636 & 3.88 & 1.63 & a & \nodata \\
\enddata

\tablenotetext{}{(This table is available in its entirety in a machine-readable form in the online journal. A portion is shown here for guidance regarding its form and content.)}
\tablenotetext{1}{k = pre-upgrade Keck/HIRES, j = post-upgrade Keck/HIRES, a = APF}
\vspace{10pt}

\label{tab:rv}
\end{deluxetable}

\section{Keplerian Analysis}
\label{sec:keplerian}

\subsection{Discovery}
\label{sec:disc}
We identify significant periodic signals in the RVs using an iterative multi-planet detection algorithm based on the two-dimensional Keplerian Lomb-Scargle (2DKLS) periodogram \citep{Otoole09}. Instead of fitting sinusoidal functions to the RV time series \citep{Lomb76, Scargle82}, we create a periodogram by fitting the RV data with Keplerian orbits at many different starting points on a 2D grid over orbital period and eccentricity. This technique allows for relative offsets and uncertainties between different data sets to be incorporated directly into the periodogram and enhances the sensitivity to moderate and high eccentricity planets.

We fit the Keplerian models using the Levenberg-Marquardt (L-M) $\chi^2$-minimization routine in the \texttt{RVLIN} IDL package \citep{WrightHoward09}. Multi-planet models are sums of single-planet models, with planet-planet gravitational interactions neglected.  Such an approximation is valid since the interaction terms are expected to be $\ll $1 \mse for non-resonant, small planets. We define a grid of search periods following the prescription of \citet{Horne86} and at each period we seed an L-M fit at five evenly-spaced eccentricity values between 0.05 and 0.7. Period and eccentricity are constrained to intervals that allow them to vary only half the distance to adjacent search periods and eccentricities. All other model parameters are free to vary, including the parameters of any previously identified planets. The period and eccentricity for previously identified planets are constrained to be within $\pm5\%$ and $^{+5}_{-10}$\%, respectively, of their initial values but all other parameters are unconstrained. This prevents slightly incorrect fits of the first detected planets from injecting periodic residuals that could mimic further planetary signals. The 2DKLS periodogram power at each point in the grid is
\begin{equation}
Z(P,e) = \frac{\chi^2 - \chi^2_{B}}{\chi^2_{B}},
\end{equation}
where $\chi^2$ is the sum of the squared residuals to the current $N+1$ planet fit, and $\chi^2_{B}$ is the sum of the squared residuals to the best $N$-planet fit. In the first iteration of the planet search (comparing a 1-planet model to a 0-planet model), $\chi^2_{B}$ is the sum of the squared error-normalized residuals to the mean (the $B$ subscript stands for baseline). The 2D periodogram is collapsed into a 1D periodogram as a function of period by taking the maximal $Z$ for each period searched (i.e. the best fit eccentricity for every period).

\begin{figure*}
         \begin{center}
              \includegraphics[width=0.85\textwidth]{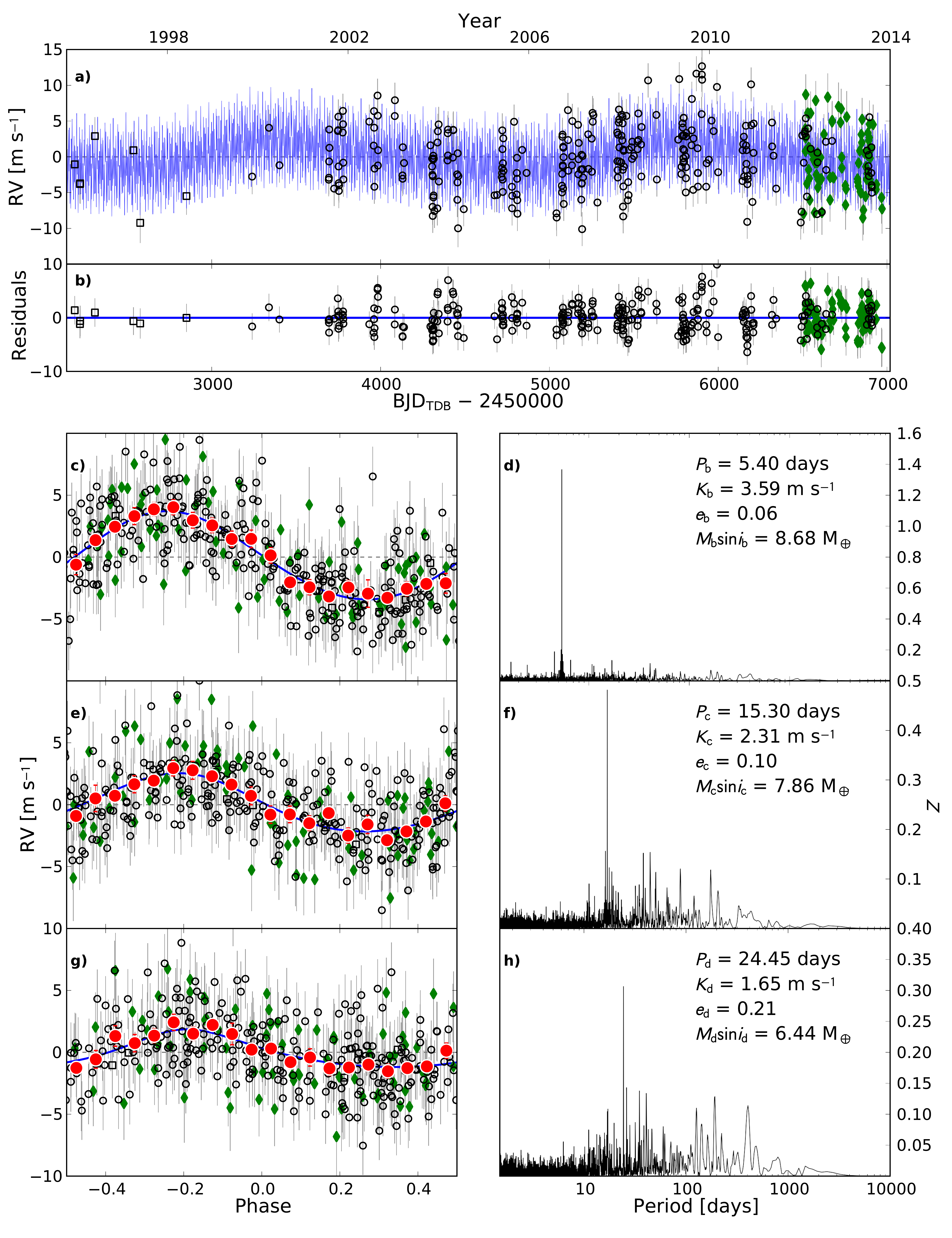}
         \end{center}
         \caption{
              Best-fit 3-planet Keplerian orbital model plus one additional long-period Keplerian to model the stellar magnetic activity cycle. The model plotted is the one that produces the lowest $\chi^2$ while the orbital parameters annotated and listed in Tables \ref{tab:params} and \ref{tab:dparams} are the median values of the posterior distributions.
              {\bf a)} Full binned RV time series. Open black squares indicate pre-upgrade Keck/HIRES data (see \S \ref{sec:keckdata}), open black circles are post-upgrade Keck/HIRES data, and filled green diamonds are APF data. The thin blue line is the best fit 3-planet plus stellar activity model. We add in quadrature the RV jitter term listed in Table \ref{tab:params} with the measurement uncertainties for all RVs.
              {\bf b)} Residuals to the best fit 3-planet plus stellar activity model.
              {\bf c)} Binned RVs phase-folded to the ephemeris of planet b. The two other planets and the long-period stellar activity signal have been subtracted. The small point colors and symbols are the same as in panel {\bf a}. For visual clarity, we also bin the velocities in 0.05 units of orbital phase (red circles). The phase-folded model for planet b is shown as the blue line.
              {\bf d)} 2DKLS periodogram comparing a 2-planet plus activity model to the full 3-planet fit when planet b is included.
              Panels {\bf e)} and {\bf f)}, and panels {\bf g)} and {\bf h)} are the same as panels {\bf c)} and {\bf d)} but for planets HD 7924 c and HD 7924 d respectively.
         }
\label{fig:fitplot}
\end{figure*}

We start the iterative planet search by comparing a 0-planet model (flat line) to a grid of 1-planet models. A strong signal at a period of \bper\ is detected at very high significance in the first iteration. This planet was initially discovered by H09 using only 22\% of the data presented in this work. When we calculate two vs.\ one planet and three vs.\ two planet periodograms, we find two additional highly significant signals with periods of \cper\ and \dper\ respectively (see Figure \ref{fig:fitplot}). These two periodic signals are best fit by Keplerian orbital models with semi-amplitudes of 2.3 \ms and 1.7 \mse, respectively, with no significant eccentricity. See Tables \ref{tab:params} and \ref{tab:dparams} for the full orbital solution.

\begin{figure}
         \begin{center}
              \includegraphics[width=0.45\textwidth]{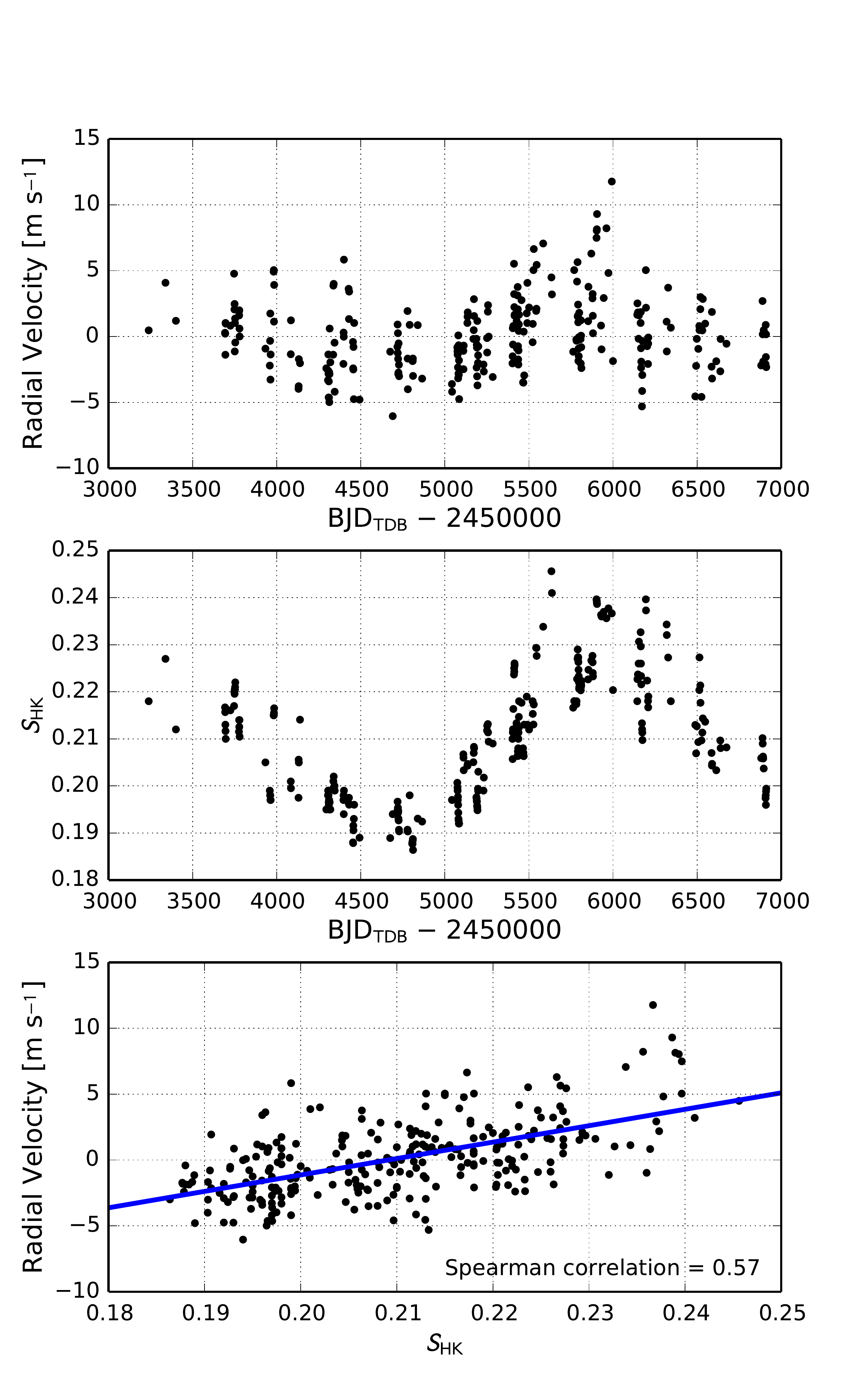}
         \end{center}
         \caption{
              Velocity-activity correlation.
              \emph{Top:} Binned RV time series of the post-upgrade Keck data with planets b, c, and d subtracted.
              \emph{Middle:} Binned \shk\ time series of the post-upgrade Keck data only. Note the similarities between the variability in the top and middle panels.
              \emph{Bottom:} Spearman rank correlation test of the velocities with \shk\ values \citep{Spearman1904}.
         }
         \label{fig:svals}
\end{figure}

We find a fourth, significant signal at \longper, but we also find a very similar signal (approximately sinusoidal with the same period and phase) upon inspection of the time series of \shk\ stellar activity measurements (see Figure \ref{fig:svals}). Although we fit for this periodic signal as an additional Keplerian we do not interpret this as an additional planet. Instead we interpret this as the signature of the stellar magnetic activity cycle. We searched  for a fifth periodic signal and found two additional marginally significant peaks at \aactper\ and \actper. Due to their marginal strengths and the fact that the two periods are related by the synodic month alias ($1/17.1\approx1/40.8+1/29.5$) we are especially cautious in the interpretation of these signals. We scrutinize these candidate signals in depth in Sections \ref{sec:alias} and \ref{sec:activity}. We also see a peak at around 40 days in a periodogram of the \shk\ time series after the long-period signal from the stellar magnetic activity cycle is removed that is presumably the signature of rotationally modulated star spots. We conclude that the 17.1 and 40.8 day signals are most likely caused by rotational modulation of starspots, but the 5.4, 15.3, and 24.5 day signals are caused by three planetary companions with minimum ($M_p\sin{i_{p}}$) masses of \bmass, \cmass, and \dmass.

\begin{deluxetable}{lrr}
    \tablecaption{Orbital Parameters}
    \tablehead{\colhead{Parameter} & \colhead{Value} & \colhead{Units}}
    \startdata
    \sidehead{\bf{Modified DE-MCMC Step Parameters}\tablenotemark{1}}
log($P_{b}$) & 0.73223 $\pm 2e-05$ & log(days)\\
$\sqrt{e_{b}}\cos{\omega_{b}}$ & 0.15 $^{+0.13}_{-0.17}$ & \\
$\sqrt{e_{b}}\sin{\omega_{b}}$ & -0.09 $^{+0.18}_{-0.15}$ & \\
log($K_{b}$) & 0.555 $^{+0.023}_{-0.024}$ & \mse\\
\vspace{5pt}\\

log($P_{c}$) & 1.184663 $^{+9.2e-05}_{-9.3e-05}$ & log(days)\\
$\sqrt{e_{c}}\cos{\omega_{c}}$ & 0.20 $^{+0.17}_{-0.24}$ & \\
$\sqrt{e_{c}}\sin{\omega_{c}}$ & 0.11 $^{+0.17}_{-0.20}$ & \\
log($K_{c}$) & 0.364 $^{+0.037}_{-0.040}$ & \ms\\
\vspace{5pt}\\

log($P_{d}$) & 1.3883 $^{+0.00027}_{-0.00031}$ & log(days)\\
$\sqrt{e_{d}}\cos{\omega_{d}}$ & 0.31 $^{+0.16}_{-0.24}$ & \\
$\sqrt{e_{d}}\sin{\omega_{d}}$ & 0.02 $^{+0.30}_{-0.37}$ & \\
log($K_{d}$) & 0.219 $^{+0.052}_{-0.057}$ & \ms\\

\hline
\sidehead{\bf{Model Parameters}}
$P_{b}$ & 5.39792 $\pm 0.00025$ & days\\
$T_{\mathrm{conj},b}$ & 2455586.38 $^{+0.086}_{-0.110}$ & \bjdtdb\\
$e_{b}$ & 0.058 $^{+0.056}_{-0.040}$ & \\
$\omega_{b}$ & 332 $^{+71}_{-50}$ & degrees\\
$K_{b}$ & 3.59 $^{+0.20}_{-0.19}$ & \ms\\
\vspace{5pt}\\

$P_{c}$ & 15.299 $^{+0.0032}_{-0.0033}$ & days\\
$T_{\mathrm{conj},c}$ & 2455586.29 $^{+0.40}_{-0.47}$ & \bjdtdb\\
$e_{c}$ & 0.098 $^{+0.096}_{-0.069}$ & \\
$\omega_{c}$ & 27 $^{+52}_{-60}$ & degrees\\
$K_{c}$ & 2.31 $^{+0.21}_{-0.20}$ & \ms\\
\vspace{5pt}\\

$P_{d}$ & 24.451 $^{+0.015}_{-0.017}$ & days\\
$T_{\mathrm{conj},d}$ & 2455579.1 $^{+1.0}_{-0.9}$ & \bjdtdb\\
$e_{d}$ & 0.21 $^{+0.13}_{-0.12}$ & \\
$\omega_{d}$ & 119 $^{+210}_{-97}$ & degrees\\
$K_{d}$ & 1.65 $\pm 0.21$ & \ms\\
$\gamma_{\rm post\mbox{-}upgrade~Keck}$ & -0.19 $\pm 0.16$ & \ms\\
$\gamma_{\rm pre\mbox{-}upgrade~Keck}$ & 2.0 $^{+1.1}_{-1.2}$ & \ms\\
$\gamma_{\rm APF}$ & 0.28 $^{+0.46}_{-0.47}$ & \ms\\
$\sigma_{\mathrm{jitt}}$ & 2.41 $^{+0.11}_{-0.10}$ & \ms\\
    \enddata
    \tablenotetext{1}{MCMC jump parameters that were modified from the physical parameters in order to speed convergence and avoid biasing parameters that must physically be finite and positive.}
    \label{tab:params}
    \end{deluxetable}

\begin{deluxetable}{lrr}
\tablewidth{0.35\textwidth}
    \tablecaption{Derived Properties}
    \tablehead{\colhead{Parameter} & \colhead{Value} & \colhead{Units}}
    \startdata
$e_b\cos{\omega_b}$ & 0.0073 $^{+0.0210}_{-0.0076}$ & \\
$e_b\sin{\omega_b}$ & -0.0031 $^{+0.0063}_{-0.0190}$ & \\
$a_{b}$ & 0.05664 $^{+0.00067}_{-0.00069}$ & AU\\
$M_{b}\sin{i_b}$ & 8.68 $^{+0.52}_{-0.51}$ & $M_{\oplus}$\\
$S_{b}$\tablenotemark{*} & 113.7 $^{+3.7}_{-3.6}$ & $S_{\oplus}$\\
$T_{eq, b}$\tablenotemark{**} & 825.9 $^{+6.6}_{-6.5}$ & K\\
\vspace{5pt}\\
$e_c\cos{\omega_c}$ & 0.017 $^{+0.051}_{-0.018}$ & \\
$e_c\sin{\omega_c}$ & 0.008 $^{+0.037}_{-0.013}$ & \\
$a_{c}$ & 0.1134 $^{+0.0013}_{-0.0014}$ & AU\\
$M_{c}\sin{i_c}$ & 7.86 $^{+0.73}_{-0.71}$ & $M_{\oplus}$\\
$S_{c}$\tablenotemark{*} & 28.35 $^{+0.92}_{-0.89}$ & $S_{\oplus}$\\
$T_{eq, c}$\tablenotemark{**} & 583.6 $^{+4.7}_{-4.6}$ & K\\
\vspace{5pt}\\
$e_d\cos{\omega_d}$ & 0.059 $^{+0.084}_{-0.054}$ & \\
$e_d\sin{\omega_d}$ & 0.001 $^{+0.076}_{-0.074}$ & \\
$a_{d}$ & 0.1551 $^{+0.0018}_{-0.0019}$ & AU\\
$M_{d}\sin{i_d}$ & 6.44 $^{+0.79}_{-0.78}$ & $M_{\oplus}$\\
$S_{d}$\tablenotemark{*} & 15.17 $^{+0.49}_{-0.48}$ & $S_{\oplus}$\\
$T_{eq, d}$\tablenotemark{**} & 499 $\pm 4$ & K\\

    \enddata
    \tablenotetext{*}{Stellar irradiance received at the planet relative to the Earth.}
    \tablenotetext{**}{Assuming a bond albedo of 0.32; the mean total albedo of super-Earth size planets \citep{Demory14}.}
    \label{tab:dparams}
    \end{deluxetable}

\subsection{Characterization}

We determine the orbital parameters and associated uncertainties for the three planet system using the \texttt{ExoPy} Differential-Evolution Markov Chain Monte Carlo \citep[DE-MCMC,][]{Braak06} engine described in \citet{Fulton13} and \citet{Knutson14}. We treat the total RV model as a sum of Keplerian orbits each parameterized by orbital period ($P_{i}$), time of inferior conjunction ($T_{\mathrm{conj}, i}$), eccentricity ($e_{i}$), argument of periastron of the star's orbit ($\omega_{i}$), and velocity semi-amplitude ($K_{i}$) where $i$ is an index corresponding to each planet (b -- d). An RV ``jitter" term ($\sigma_{\mathrm{jitt}}$) is
added in quadrature with the measurement uncertainties at each step in the MCMC chains. We also fit for independent RV zero-points for the APF, pre-upgrade Keck, and post-upgrade Keck data. The long-period RV signal presumably caused by the stellar magnetic activity cycle is treated as an additional Keplerian orbit with the same free parameters as for each of the three planets. In order to speed convergence and avoid biasing parameters that must physically be finite and positive we step in the transformed and/or combinations of parameters listed in Table \ref{tab:params}. We add a $\chi^2$ penalty for large jitter values of the following form

\begin{equation}
\chi^2_{\rm new} = \chi^2 + 2\sum_n \ln{\sqrt{2\pi(\sigma_{\text{vel}, n}^2 + \sigma_{\mathrm{jitt}}^2)}},
\end{equation}
where $\sigma_{\text{vel}, n}$ is the velocity uncertainty for each of the $n$ measurements \citep{Johnson11}.

We fit for the 23 free parameters by running 46 chains in parallel continuously checking for convergence using the prescription of \citet{Eastman13}. When the number of independent draws \citep[$T_{z}$ as defined by][]{Ford06} is greater than 1000 and the Gelman-Rubin statistic \citep{Gelman03, Holman06} is within 1\% of unity for all free step parameters we halt the fitting process and compile the results in Table \ref{tab:params}. We also list additional derived properties of the system in Table \ref{tab:dparams} that depend on the stellar properties listed in Table \ref{tab:stellar_params}.

\subsection{False Alarm Assessment}
\label{sec:fap}
\begin{figure}
\centering
\includegraphics[width=0.45\textwidth]{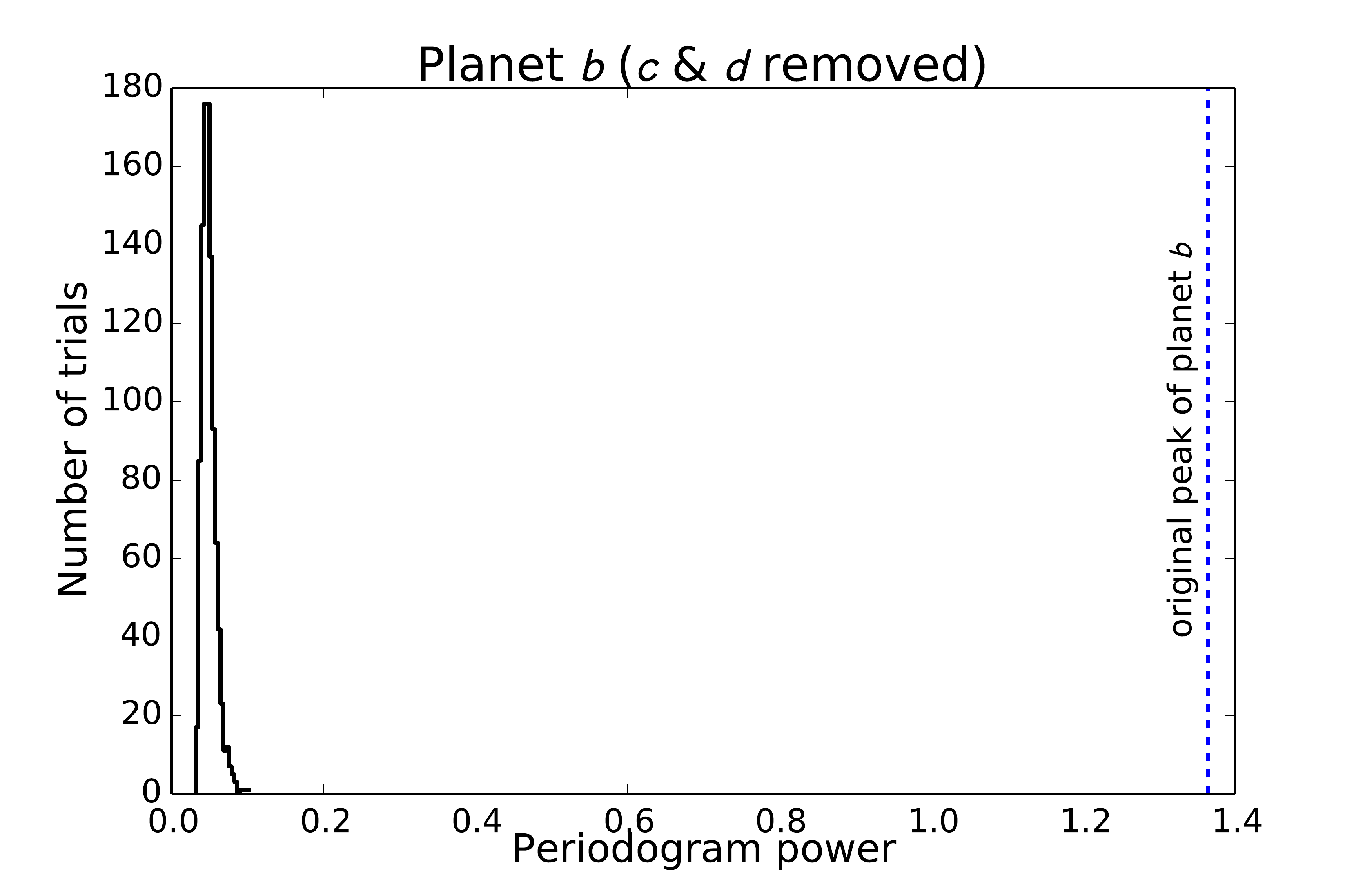}
\includegraphics[width=0.45\textwidth]{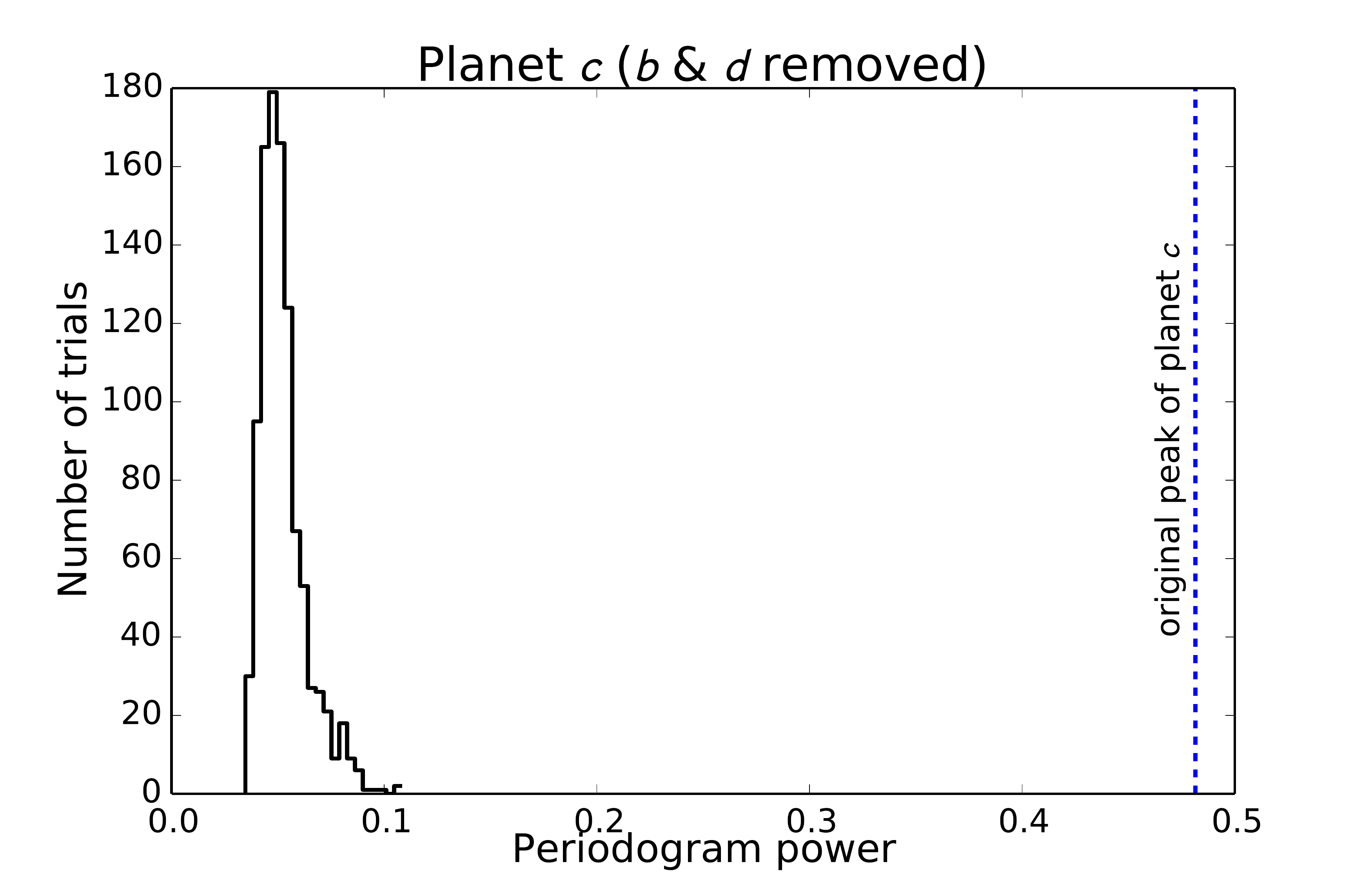}
\includegraphics[width=0.45\textwidth]{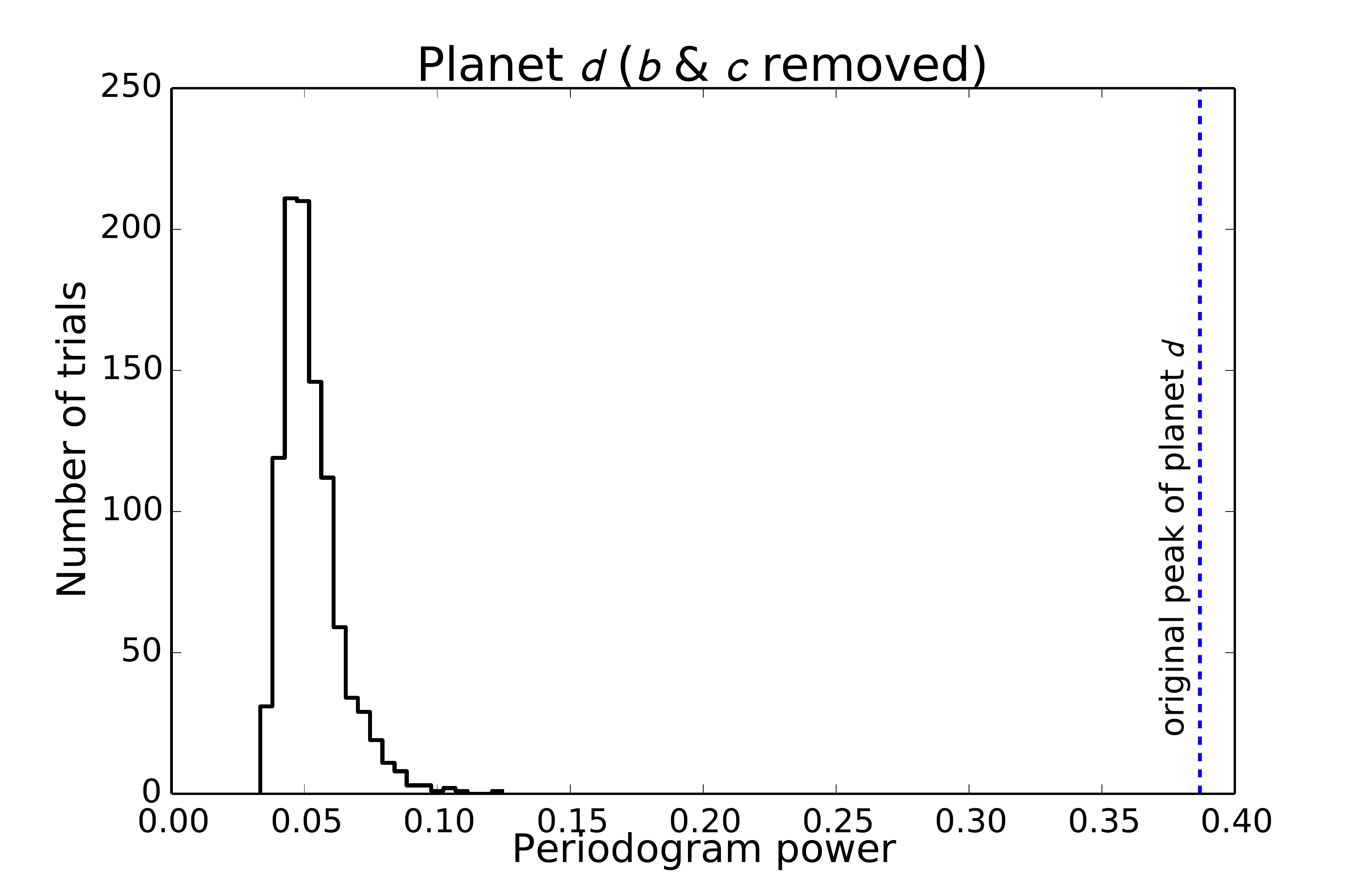}
\caption{
              \emph{Top:} Distribution of maximum periodogram peak heights for 1000 2DKLS periodograms of scrambled RV time series. For each periodogram planets c, d, and the magnetic activity cycle were subtracted before scrambling the data set. The vertical dashed blue line marks the height of the original peak for planet b which is clearly separated from the distribution of peaks caused by random fluctuations.
              \emph{Middle:} Same as the top panel for planet c.
              \emph{Bottom:} Same as the top panel for planet d.
}
\label{fig:boot}
\end{figure}

We attempted to empirically determine the probability that Gaussian random noise in the data could conspire to produce an apparent periodic signal with similar significance to the periodogram peaks corresponding to each of the planets. We calculated 1000 2DKLS periodograms, each time scrambling the velocities in a random order drawn from a uniform distribution. We located and measured the height of the global maxima of each periodogram and compare the distribution of these maxima to the periodogram peak heights at the periods of the three planets in the original periodograms. The power for each periodogram within the set of 1000 is a $\Delta\chi^2$ between a 2-planet plus activity model to a 3-planet model assuming the other two planets as ``known". Figure \ref{fig:boot} shows that the distribution of maxima from the periodograms of the scrambled data are clearly separated from the original peaks. None of the trials produce periodogram peaks anywhere near the heights of the original peaks corresponding to the three planets. Because the 2DKLS periodogram allows eccentric solutions, we explore the scrambled RVs for non-sinusoidal solutions and are therefore sensitive to a wide variety of false alarm signals. We conclude that the false alarm probabilities for all three planets are $<0.001$.

\subsection{Searching for Period Aliases}
\label{sec:alias}

The high cadence of the APF data set allows us to explore short-period orbital solutions. Traditional observations on large telescopes such as Keck often yield only a few nights of data per year, making it difficult to determine whether a short orbital period is an alias of a longer period, or a true physical signal. Although planets with orbital periods shorter than one day are uncommon in the galaxy \citep[around $0.83\pm0.18$ \% of K dwarfs;][]{Sanchis-Ojeda14}, eleven ultra-short period planets have been found\footnote{Based on a 2014 Nov 20 query of exoplanets.org \citep{Wright11, Han14}} due to their high detectability. To search as carefully as possible for short period signals, we use the unbinned data sets, which consist of \numkeck\ RVs from Keck and \numapf\ RVs from the APF.  Our use of the unbinned data in this section also provides independent confirmation of the results obtained with the binned data above.

\citet{Dawson10} outline a rigorous procedure to distinguish between physical and alias periods. Our method for finding the orbital periods and distinguishing aliases is as follows:
\begin{enumerate}
\item Determine the window function of the data to understand which aliases are likely to appear.
\item Compute the periodogram of the data, determining the power and phase at each input frequency.
\item If there is a strong peak in the periodogram, fit an N-planet Keplerian (starting with N=1), using the periodogram peak as the trial period.
\item Subtract the N-planet Keplerian from the data.
\item Compute the periodogram of the Nth planet in the model and compare it to the periodogram of the Nth planet in the data minus the model of the other planets.
\item If a second peak in the periodogram has similar height to the tallest peak and is located at an alias period, repeat steps 3-5 using that trial period.
\item If you explored an alias period, choose the model that minimizes $\chi^2$ and best reproduces the observed periodogram.  Subtract this model from the RVs.
\item Treat the residuals as the new data set and go back to step 2.  Examine the residuals from the N-planet fit for additional planets, and continue until there are no more signals in the periodogram.
\end{enumerate}

\begin{figure}[htbp] 
   \centering
\includegraphics[width=3in]{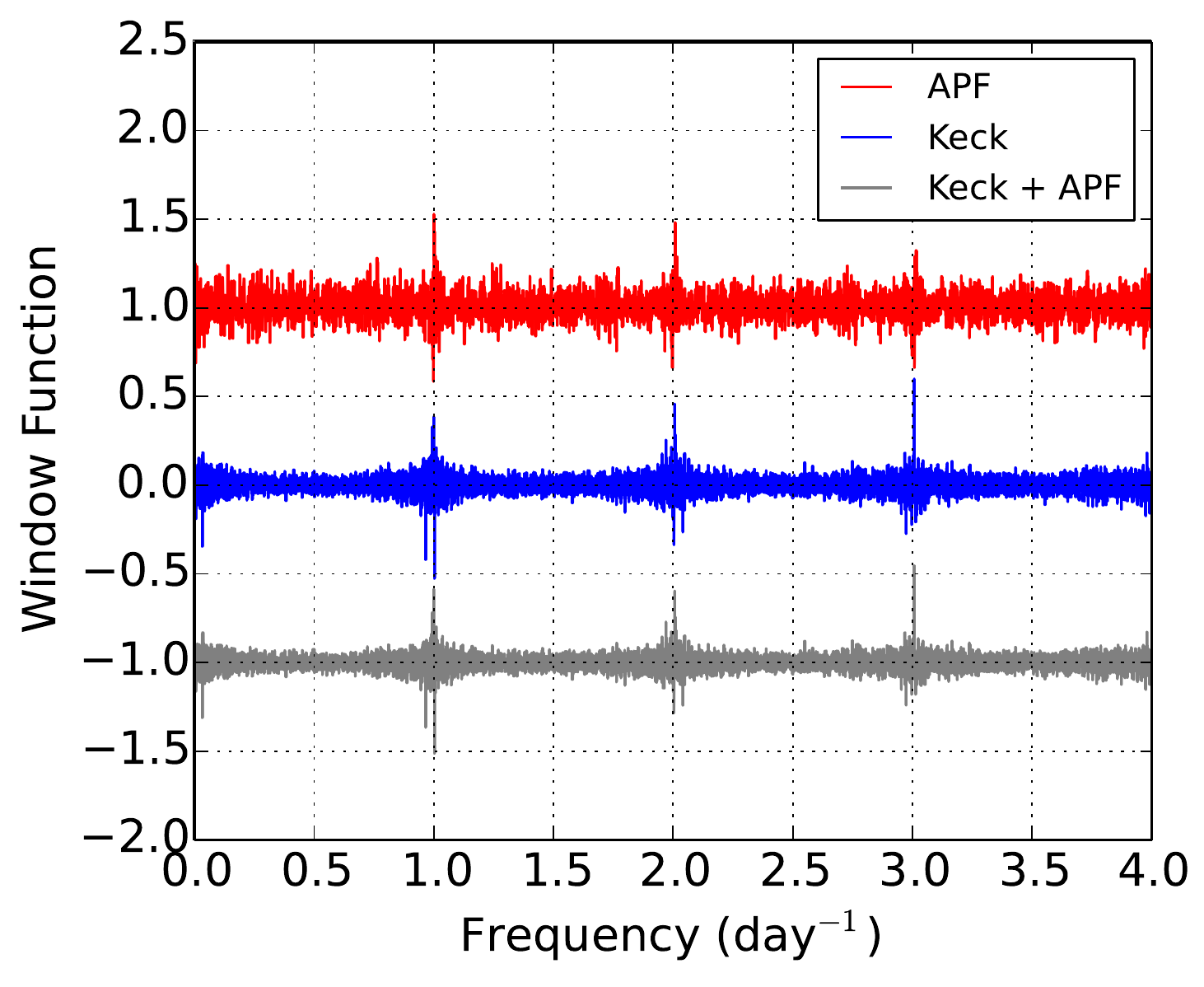}
   \caption{The window function of the Keck and APF RV time series.  While the APF window function has some power at frequency multiples of one day, it is flat otherwise, whereas the Keck window function has power at low frequencies (corresponding to long-period aliases) and power in broader swaths around the frequency multiples of one day.}
   \label{fig:window}
\end{figure}

The window functions of the individual and combined Keck and APF RV time series are shown in Figure \ref{fig:window}.  The window function is given by
\begin{equation}
W(\nu) = \frac{1}{N}\sum\limits_{j=1}^N \exp(-2\pi i \nu t_j),
\end{equation}
where $\nu$ is the frequency in units of days$^{-1}$ and $t_{j}$ is the time of the $j$th observation.  The data sets are complementary: 109 RVs from APF over the past year are well-distributed over the months and the year, and so are only susceptible to the daily aliases, whereas the \numkeck\ RVs from Keck over the last decade are distributed in a way that gives some power to daily aliases as well as longer-period aliases. The power in the combined window function illustrates that we might be susceptible to daily aliases, and weak signals in the periodogram might even be susceptible to monthly or yearly aliases.

\begin{figure*}[htbp] 
   \centering
\includegraphics[width=6in]{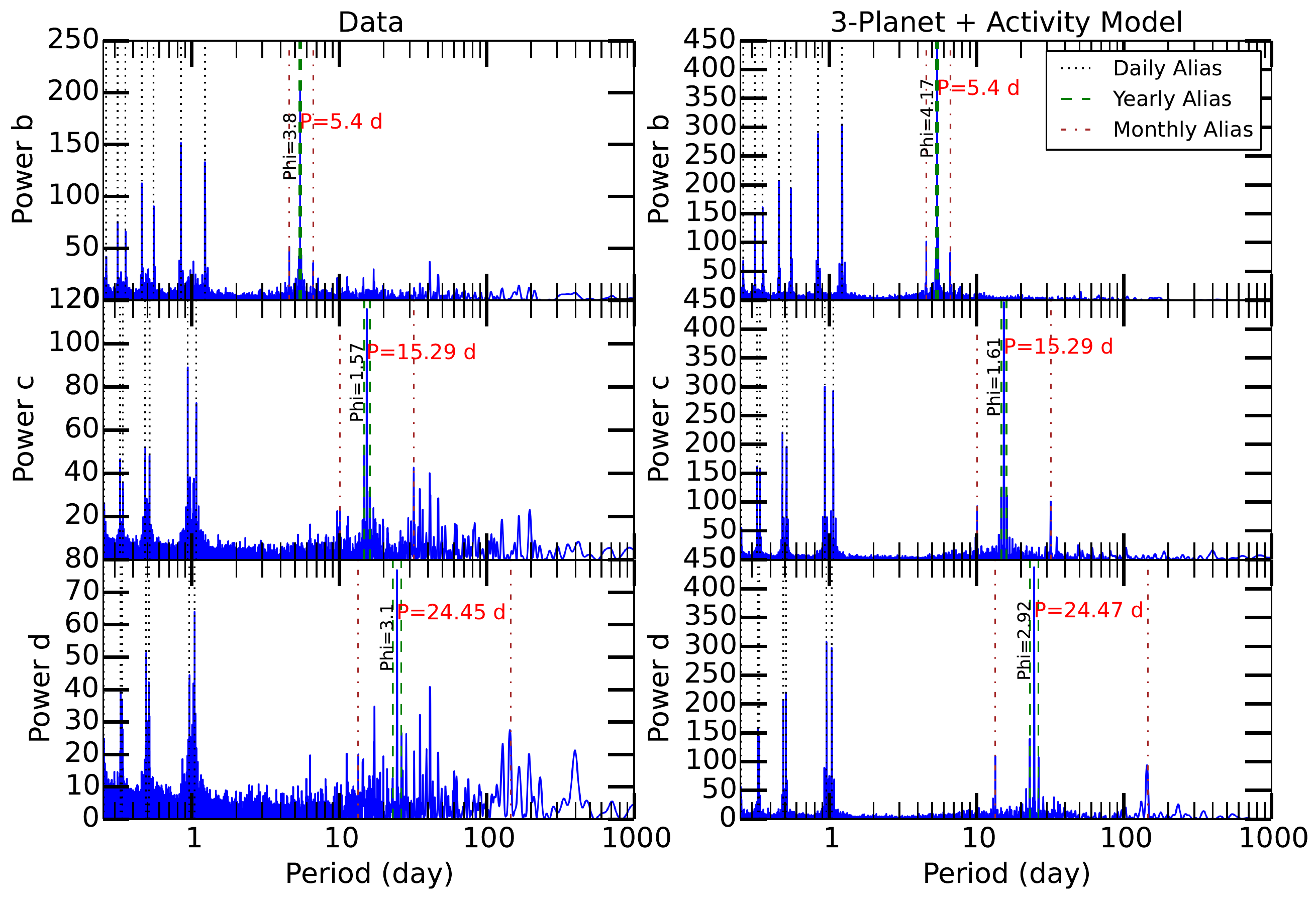}
   \caption{Left: LS periodograms of the data associated with each planet identified, from top to bottom: planet b, planet c, and planet d.  In each panel, signals from the other planets and stellar activity have been subtracted.  The phase of the frequency associated with the peak is given in radians.  Right: LS periodograms of the best Keplerian model for each of the planets, from top to bottom, planets b, c, and d.  The periodogram of each Keplerian model reproduces the peak period and alias structure of the data.}
\label{fig:data+model}
   \end{figure*}

\begin{figure*}[htbp] 
   \centering
 \includegraphics[width=6in]{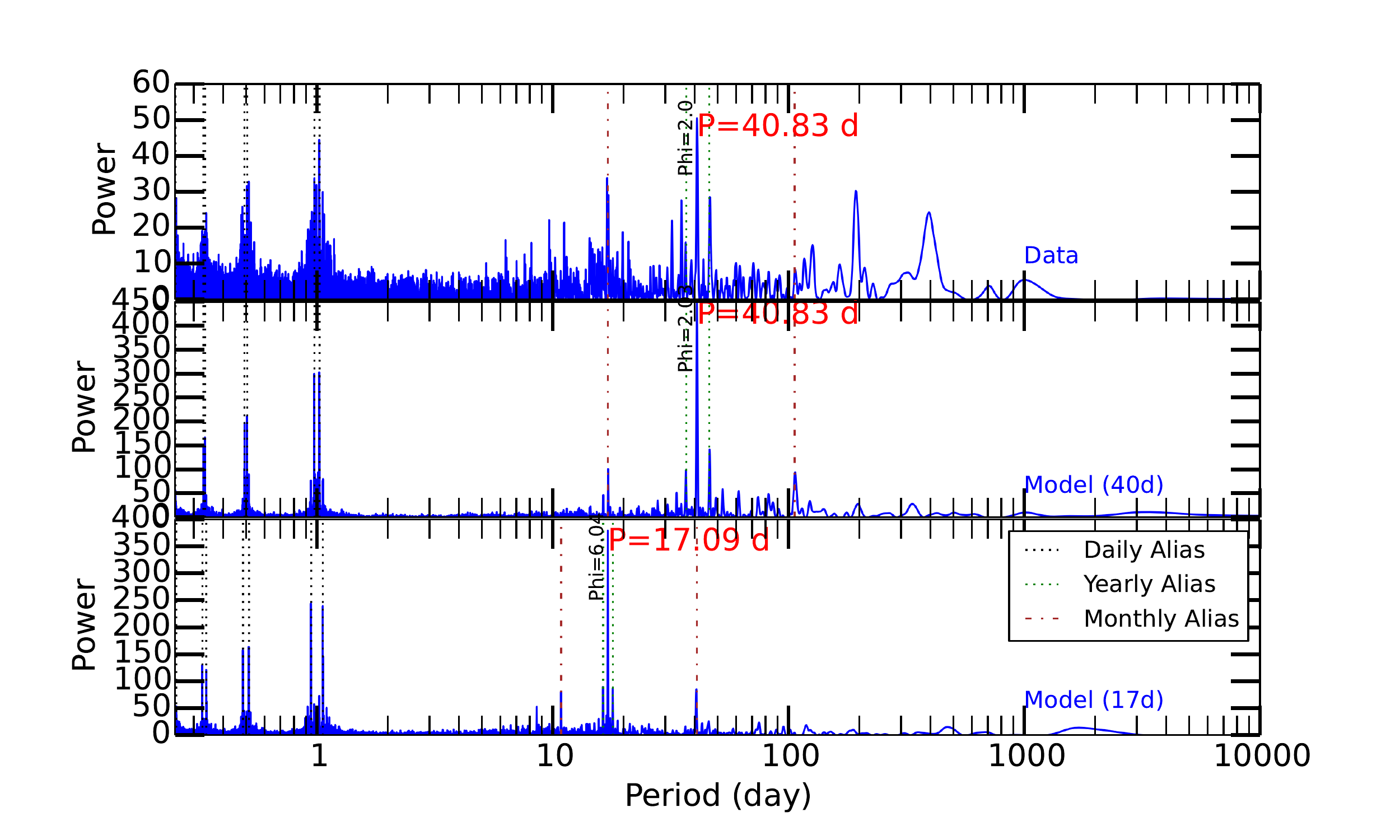}
   \caption{Top: a candidate periodic signal at 40.8 or 17.1 days emerges in the periodogram of the residuals to the 3-planet plus long-term stellar activity model. These two periods are related to each other by the one synodic month alias. Either 40.8 days or 17.1 days could correspond to the rotation period of the star, although the 40.8 day period is more prominent in the periodogram of the \shk\ values.  To test both periods, we model the best-fit Keplerian at 40.83 days (center) and 17.1 days (bottom) and show their periodograms, complete with phase information.}
   \label{fig:40/17}
\end{figure*}

To take the periodogram of the time series, we use a version of \texttt{fasper} \citep{Press89} written for Python.  We find the same peaks in the periodogram of the data and residuals at 5.4, 15.3, and \dper\ that we interpret as planets.  The periodograms of the data and periodograms of the Keplerian models associated with these periods are shown in Figure \ref{fig:data+model}.  Again, we recover a fourth peak at 2570 days that we attribute to long-term stellar activity due to the correlation between the RVs and the \shk\ values. The 3 planet plus stellar activity model found with the alias-search method yields periods, eccentricities, and planet masses within 1-$\sigma$ of the results quoted in Table \ref{tab:params} for all three planets. $\omega_b$ is consistent within 1-$\sigma$, but inconsistent for planets c and d. However, the arguments of periastron for all three planets are poorly constrained due to their nearly circular orbits. We find a fifth peak at \actper, which is also prominent in the periodogram of the stellar activity and is likely the rotation period of the star.  The 40.8-day signal has a strong alias at \aactper\, and so we test Keplerian models at both \actper\ and \aactper\ to discriminate which is the true signal and which is the alias (see Figure \ref{fig:40/17}).  We find that the periodogram of the model 40.8 day signal better matches the alias structure in the periodogram of the data, and so we prefer \actper\ as a candidate stellar rotation period.


\subsection{Chromospheric Activity}
\label{sec:activity}

Although this star is relatively inactive with \rphk\ values in the literature between $-4.89$ \citep{Isaacson10} and $-4.85$ \citep{CantoMartins11}, some low-level chromospheric activity is detectable with our high-precision RVs. The most obvious feature is the long period signal with a period of $\approx$6.6 years and an amplitude of $\approx$5 \ms. Although this signal looks promising as a long-period sub-Jupiter mass planet candidate, upon inspection of the \shk\ time series we notice that this activity indicator is highly correlated with this long-period signal in the velocities (see Figure \ref{fig:svals}). We interpret this signal as the signature of the stellar magnetic activity cycle of which we have observed nearly two full cycles. If we subtract this long-period signal from the \shk\ values and make a periodogram of the residuals we find a marginally significant peak at $\approx$41 days. This is most likely caused by rotational modulation of starspots, because 41 days is near the expected rotation period \citep[38 days,][]{Isaacson10} for a star of this spectral type and age.

To gain confidence in the source of these Doppler signals (activity or planets) we employed a time-dependent RV-\shk\ decorrelation that probes evolution of the RV periodogram on timescales of the stellar rotation.
Simple linear decorrelation is limited by the natural phase offset between RV and activity signals. At any given time, the measured RV and activity both depend on the flux-weighted fractional coverage of the magnetically active region on the stellar disk.  The measured \shk\ peaks when the magnetically active regions are closest to the disk center, where its projected area is largest and the star would otherwise be brightest. However, the RV approaches zero at the center of the stellar disk. The activity-induced RV shift is maximized when the active region is closer to the limb, $\sim$1/8 phase from center \citep[e.g.][]{Queloz01, Aigrain12}. In addition, the magnetically active regions continuously evolve on timescales of several stellar rotation periods weakening the RV-\shk\ correlations over long timescales. 

Motivated by these limitations, our alternative RV-\shk\ decorrelation method is effective on timescales of both the long-term magnetic activity cycle and the stellar rotation period.  All APF data are excluded from this analysis because activity measurements are not presently available. 

Starting with the same outlier-removed time-series as described above, we bin together any radial velocity and \shk\ measurements taken on the same night.  The period of the long-term stellar magnetic activity cycle is then identified from a Lomb-Scargle (LS) periodogram of the activity time-series.   That period serves as the initial guess in a subsequent single-Keplerian fit to the RV time-series using \texttt{rv\_mp\_fit}.   Next, planet candidates are identified using an iterative approach in which the number of fitted planets $N$ increases by one starting at $N=1$.  $N+1$ Keplerians are fit at each step, one accounting for the long-term stellar magnetic activity cycle:  
\begin{enumerate}
	\begin{item}
		\textbf{Determine period of $N_{\mathrm{th}}$ planet candidate:} Identify the period, $P_N$, of the highest peak in the LS periodogram of the residuals to the $N$-planet Keplerian fit to the RV time-series.  
	\end{item}
	\begin{item}
		\textbf{Obtain best $N$-planet solution}: Perform an $(N+1)$-planet Keplerian fit with initial guesses being the $N$ Keplerians from the previous fit plus a Keplerian with initial period guess $P_N$. 
	\end{item}
	\begin{item}
		\textbf{Repeat}: Steps 1-2 for $N = N + 1$.   
	\end{item}
\end{enumerate}

We wish to subsequently remove residual RV signatures that can be confidently attributed to magnetically active regions rotating around the stellar surface.  Since the lifespan of such regions is typically a few rotation periods, each measured RV was decorrelated with \shk\ activity measurements within $\pm$100 days. We chose $\pm$100 day windows because such windows were long enough to allow us to detect the 41 day rotation period in each window, and short enough to capture long-term variability in the power spectrum on time-scales of hundreds of days. However, before discussing our decorrelation technique, we investigate the degree to which our RV and \shk\ measurements are correlated and the reliability of our planet candidate signals. 

\begin{figure}
   \includegraphics[scale=0.4]{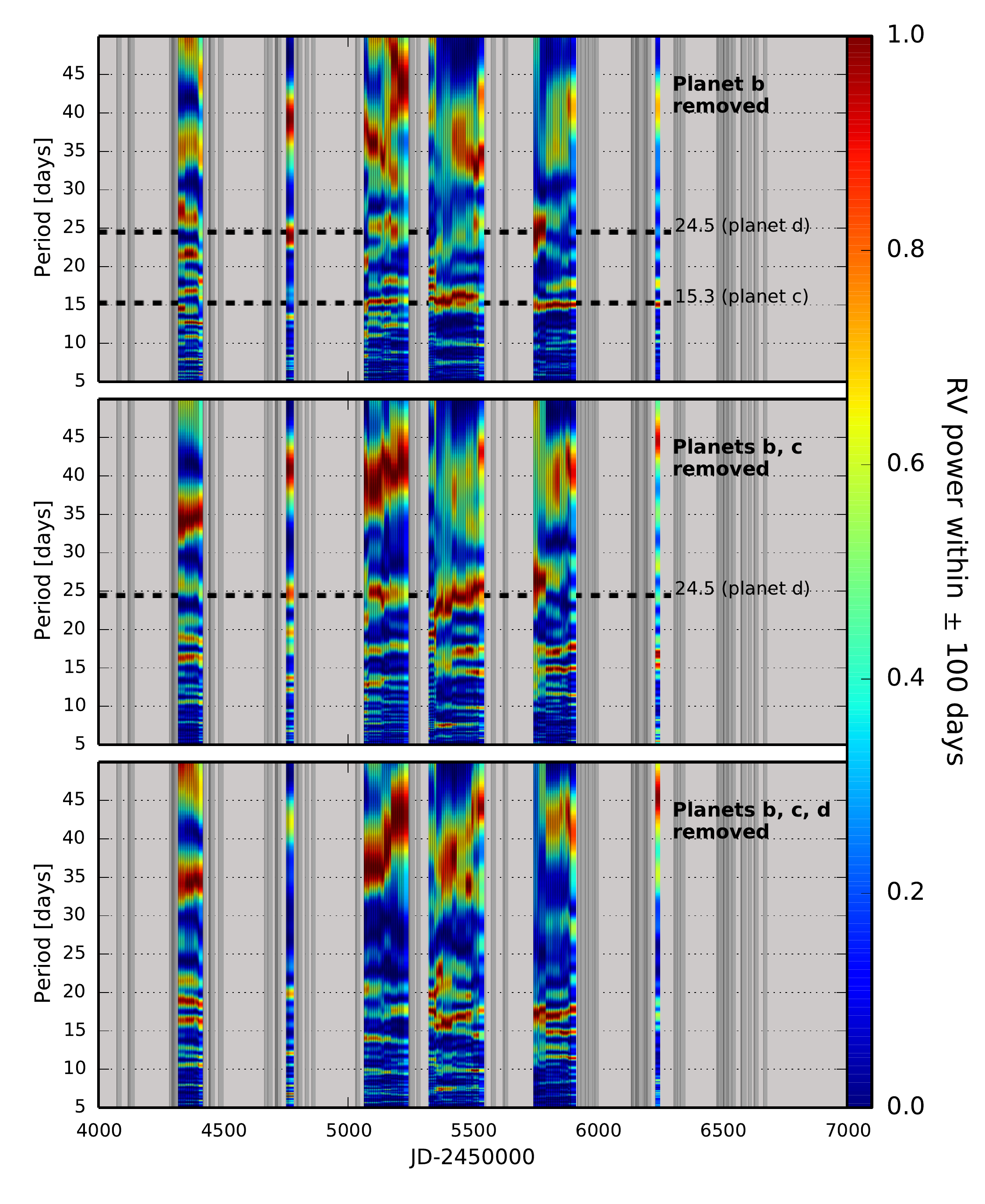}
   \caption{\emph{Top}:  Periodogram of Keck HIRES RV measurements $\pm$ 100-days versus time (JD-2450000) with our Keplerian model of planet b (P=\bper) removed.  Each colored, single-pixel column is a unique periodogram of all RV measurements within a 200-day window centered on the time indicated on the horizontal axis.  Dark grey boxes indicate times of observation having fewer than 24 measurements of both RV and \shk\ within $\pm$ 100 days.  Each periodogram has been normalized independently by dividing all periodogram powers by the maximum periodgram value.  \emph{Middle}:  Same as top, with planets b and c removed. \emph{Bottom}:  Same as top, with planets b, c, and d removed.} 
   \label{fig:RVdecorr}
\end{figure}

Figure \ref{fig:RVdecorr} shows a time-series of LS periodograms corresponding to all RVs measured within $\pm$100 days.  Each panel contains more than 100 individual columns corresponding to each periodogram derived from all data within $\pm$100 days of the times for each column.  These periodograms are only displayed at times when a minimum data cadence requirement (discussed below) is met and are grayed out elsewhere.  Individual periodograms have been normalized independently to facilitate inter-epoch comparison of the relative distribution of power across all periods.   In each panel from top to bottom, an additional planet signal has been removed from the data. The long-term magnetic activity cycle is removed from the data in all three panels.  The fitted longterm magnetic activity-induced component has also been removed in all cases.  After removing planet b (top panel), the majority of epochs/columns show significant power at \cper, corresponding to planet c.  The few exceptions are epochs with the lowest cadence.  The higher cadence epochs also have significant power near \dper, corresponding to planet d.  The 24.5-day signal is more apparent after planet c is also removed (middle panel).   The presence of 15.3-day and 24.5-day peaks across all high-cadence epochs is consistent with the coherent RV signature of a planet.  

Figure \ref{fig:RVdecorr} also reveals significant RV power at $\sim$30--50-day periods which can be attributed to magnetically active regions (i.e. spots and plages) rotating around the stellar surface at the $\sim$40.8-day stellar rotation period.  The activity-induced RV shifts dominate the periodograms after the planets b, c and d have been removed (bottom panel).  A number of epochs also show substantial power at $\sim$ 17.1 days.  This is unlikely to be the signature of a fourth planet as the period corresponds to the monthly synodic alias of the 40.8-day rotation period.   Moreover, a system of planets with 15.3-day and 17.1-day periods is unlikely to be dynamically stable. 
  
\begin{figure}
   \includegraphics[scale=0.4]{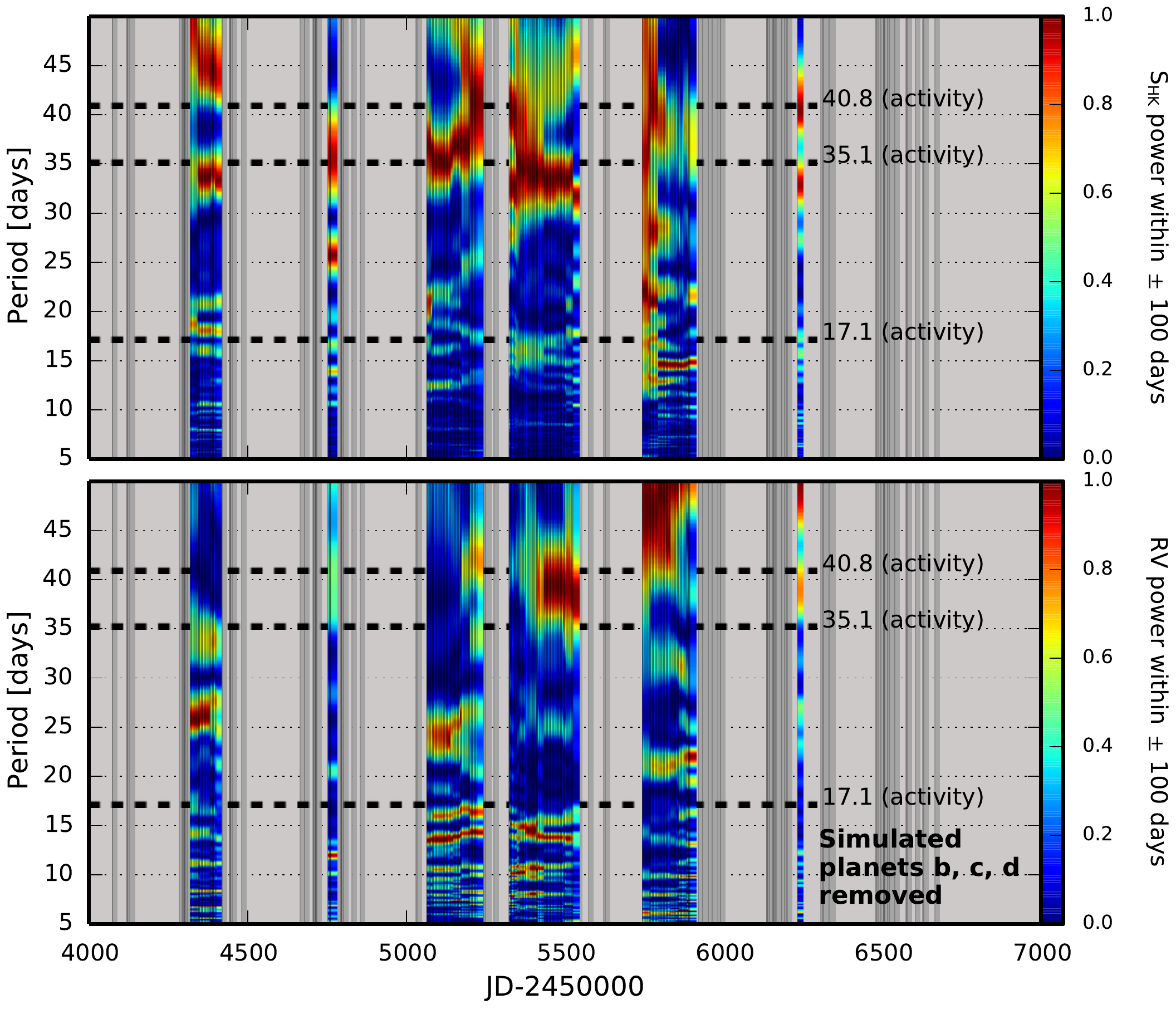}
   \caption{\emph{Top}:  Same as Figure \ref{fig:RVdecorr} but for \shk\ (stellar activity) data.  \emph{Bottom}:  Same as Figure \ref{fig:RVdecorr} bottom panel but for a synthetic RV time-series created by superposing a 1.46 \mse, 40.8-day sinusoid with the best 3-planet model to the real Keck RVs and 2.36 \ms Gaussian noise, then fitting for and removing the best 3-planet model (P=5.4, 15.3, 24.5 days).   Time sampling matches the real-data.  Dark grey boxes indicate times of observation having fewer than 24 measurements of both RV and \shk\ within $\pm$ 100 days.  Epoch-to-epoch fluctuation in the recovered period of the injected 40.8-day signal is an artifact of time-sampling, Keplerian fitting noise, and injected Gaussian noise.}
   \label{fig:shk}
\end{figure}

The direct correlation between RV and stellar activity on stellar rotation timescales is further highlighted in Figure \ref{fig:shk}.  The top panel is the \shk-analog of Figure \ref{fig:RVdecorr} in that each column is the running periodogram of all \shk\ measurements within $\pm$ 100 days.  The suite of 30-50-day peaks mirror those in the RV periodograms of Figure \ref{fig:RVdecorr}, indicating that magnetic activity is responsible for RV variation on these timescales.    

The observed $\sim$20-day range in the period of these activity-induced peaks might seem surprisingly large, given that differential stellar rotation is likely of order several days (e.g. the observed difference in rotation period at the Sun's equator and 60$^{\circ}$ latitude is $\sim$5 days).   However,  some of this variation could also be statistical noise, simply an artifact of the limited sampling of $\sim$40-day signals over a relatively small time baseline.  To test this hypothesis, we generated a synthetic RV time series of the exact same cadence as our real Keck observations and attempted to recover an injected signal of period \actper.  We first injected our three-planet model, adding Gaussian noise with standard deviation 2.36 \mse, the median velocity error in the post-upgrade Keck dataset (see Table \ref{tab:performance}).  We then superposed the 40.8-day sinusoid of semi-amplitude 1.46 \mse.  The chosen semi-amplitude is the median of semi-amplitudes of all Keplerian fits to activity-induced RVs as determined by our decorrelation algorithm (see below).   We applied the same planet-detection algorithm described above to these simulated data and subtracted three Keplerians from the best 4-Keplerian model, leaving only the 40.8-day signal.  

The bottom plot in Figure \ref{fig:shk} shows the resulting 200-day-running-window periodogram, akin to the bottom plot in Figure \ref{fig:RVdecorr}.   The $\sim$10--15 day fluctuation in the recovered period of the injected 40.8-day signal is consistent with activity-induced features in our real \shk\ and RV data. We conclude that the varying power of the RV signal with periods of 30-50 days observed in the real data is a limitation of the observing cadence and not intrinsic to the star.

The strong correlation observed between RV and stellar activity indicates tremendous potential for decorrelation to validate our planet candidates.   We apply a decorrelation algorithm that iterates through each RV measurement $v(t_i)$ where $t_i$ is the time of the $i_{\mathrm{th}}$ measurement:

\begin{enumerate}
	\begin{item}
		Define subsets $S_i$ and $V_i$, consisting of all \shk\ and RV measurements within 100 days of time $t_i$.  
	\end{item}
	\begin{item}
	         \label{itm:Nmin}
		Define $n_{\mathrm{S}}$ and $n_{\mathrm{V}}$ as the number of datapoints in $S_i$ and $V_i$ respectively.  If $n_{S}$ and $n_{V} > n_{\mathrm{min}}$ then proceed to next step, otherwise revert to step 1 for $i = i + 1$.
	\end{item}
	\begin{item}
		Identify the period, $P_i$, of the highest peak in the LS periodogram of $S_i$.  
	\end{item}
	\begin{item}
        		Perform a single-Keplerian fit to $V_i$ with fixed period $P_i$.
 	\end{item}
	\begin{item}
        		Subtract Keplerian fit from $v(t_i)$.  
 	\end{item}
	\begin{item}
        		Repeat for $i=i + 1$.  
 	\end{item}
\end{enumerate}

As step \ref{itm:Nmin} indicates, we only decorrelated an RV measurement with \shk\ activity if both the number of \shk\ and RV measurements within $\pm$100 days, $n_{S}$ and $n_{V}$ respectively, exceeded some minimum number, $n_{\mathrm{min}}$.  This was done to avoid removal of spurious correlations.  $n_{\mathrm{min}}$ was chosen by examining the dependence of correlation significance on $n_{S}$ and $n_{V}$.   More specifically, for each RV measurement, we computed the Pearson product-moment correlation coefficient, $r$, between RV and \shk\ periodograms from all data within $\pm$100 days.  In cases of $n_{S}$ and $n_{V}$ between $\sim$10--20, $r$ values were widely distributed between 0-1, suggesting correlations were spurious.  Supporting this notion, the best-fitting Keplerians to the activity-correlated RV noise had unusually large semi-amplitudes, as high as 10 \mse, even in cases where $r > 0.5$.   In contrast, when $n_{S}$ and $n_{V}$ exceeded 23, $r$ ranged from 0.34-0.76, with a median of 0.51 and semi-amplitudes were a more reasonable 1-3 \mse.  We therefore adopted $n_{\mathrm{min}}$ = 23 resulting in activity decorrelation of 115 of \binkeck\ RV measurements.

\begin{figure}
   \includegraphics[scale=0.4]{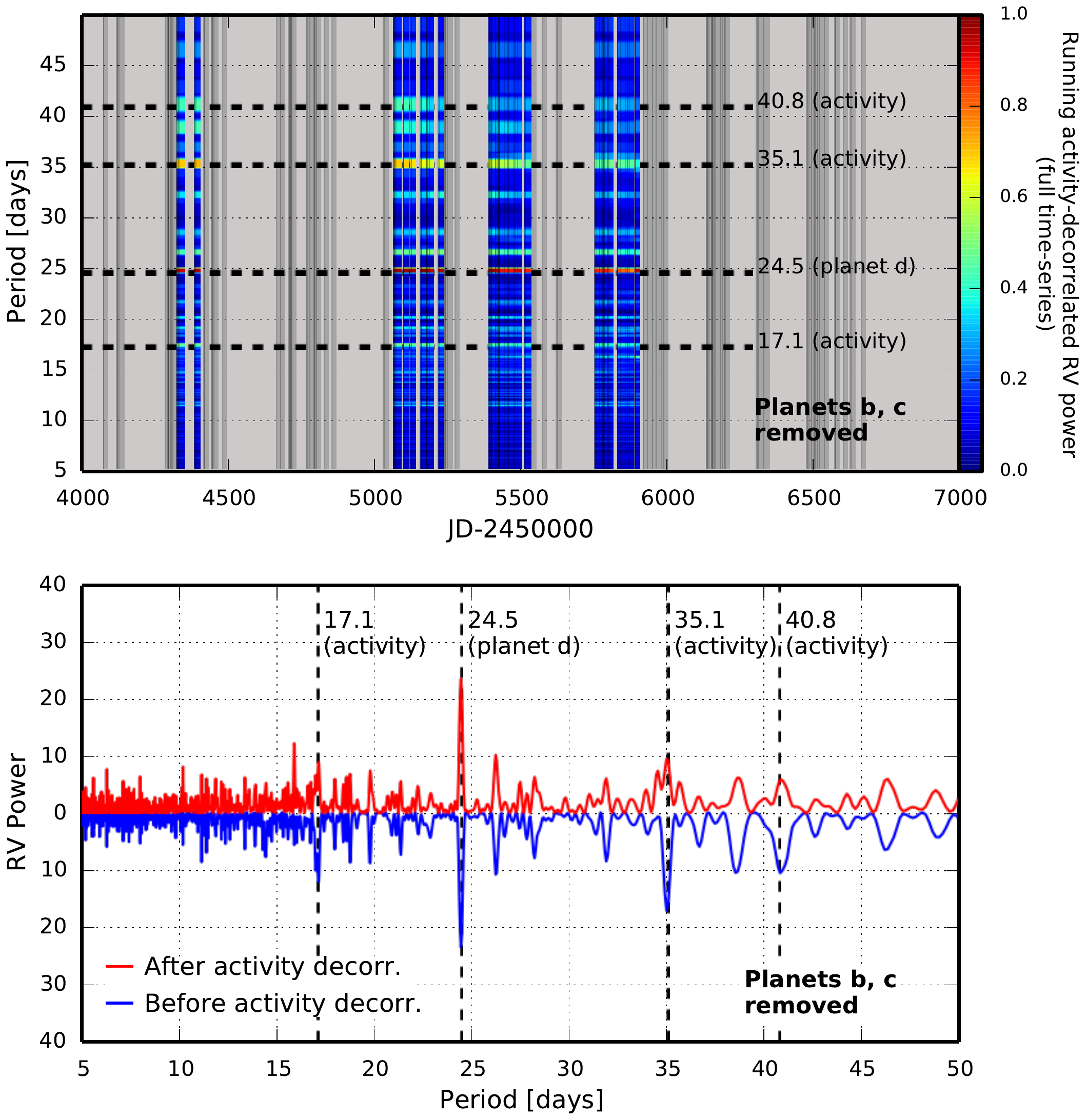}
   \caption{\emph{Top}:  Running periodogram of the entire RV time-series, where only the subset of RV measurements that precede each time have been decorrelated with stellar activity.  Keplerian models of planets b (P=5.4 days), c (P=15.3 days) and the long-term-stellar-magnetic-activity have been removed.  Dark grey boxes indicate times of observation having fewer than 24 measurements of both RV and \shk\ within $\pm$ 100 days. All periodogram powers have been normalized to the same color scale. \emph{Bottom}:  Periodogram of entire RV time-series before and after decorrelation with stellar activity. The periods of peaks attributed to planets and stellar activity are labeled accordingly by dashed vertical lines.}
   \label{fig:RVall}
\end{figure}

The top panel of Figure \ref{fig:RVall}, shows the evolution of the periodogram of the entire RV time series as the activity decorrelation scheme is cumulatively applied to data points in chronological order.  That is, the columnar periodogram at each given time, $t$, corresponds to that of the entire RV time series, with activity decorrelation of all RV measurements before and including time $t$.   Planets b and c have been removed as well as the long term magnetic activity.  The bottom panel dispays the RV periodograms before and after activity decorrelation, corresponding to the leftmost and rightmost periodograms in the top panel respectively.   After decorrelation the periodogram has significant power at 24.5, 35.1, and \actper\ and several other more moderate peaks.  However, after decorrelation (rightmost column), the 24.5 signal dominates, in support of its planetary origin, while most other peaks, including those at \actper\ and 35.1 days, have been drastically reduced, consistent with manifestations of stellar activity.  The single exception is the peak at \aactper.  While it is unclear why this peak does not vanish, we remind the reader that is the monthly synodic alias of the 40.8 day stellar rotation period, and a planet with such an orbital period would likely be dynamically unstable with planet c (P=15.3 days).  The 3-planet result of this analysis is consistent with the other analyses previously described. This method of decorrelating stellar noise from RV measurements will be useful for distinguishing planets from stellar activity in other high-cadence RV planet searches.

\section{Photometry}
Photometric observations of planet host stars can be used to measure stellar rotation, detect planetary transits, and sense false planet detections in RV signals. We collected 1855 differential photometry measurements of HD 7924 using the T8 0.8 m automated photometric telescope (APT) at Fairborn observatory in Arizona. The measurements were collected between 31 December, 2006 and 1 December, 2014. The instrument uses two photomultiplier tubes to measure flux in Str\"{o}mgren $b$ and $y$ filters simultaneously. The $b$ and $y$ measurements are later combined into a single $(b+y)/2$ passband to improve signal to noise. The telescope nods between the target star and several comparison stars.  A more detailed description of the observing procedure and data reduction can be found in H09 and \citet{Henry13}.
We noticed season to season offsets in the photometry of HD 7924 with a maximum amplitude of $\sim1.5$ mmag. These offsets are well correlated with the \shk\ values during the same time. To better characterize the photometric variability on short timescales we remove these offsets for the subsequent analysis.

\begin{figure}
         \begin{center}
              \includegraphics[width=0.45\textwidth]{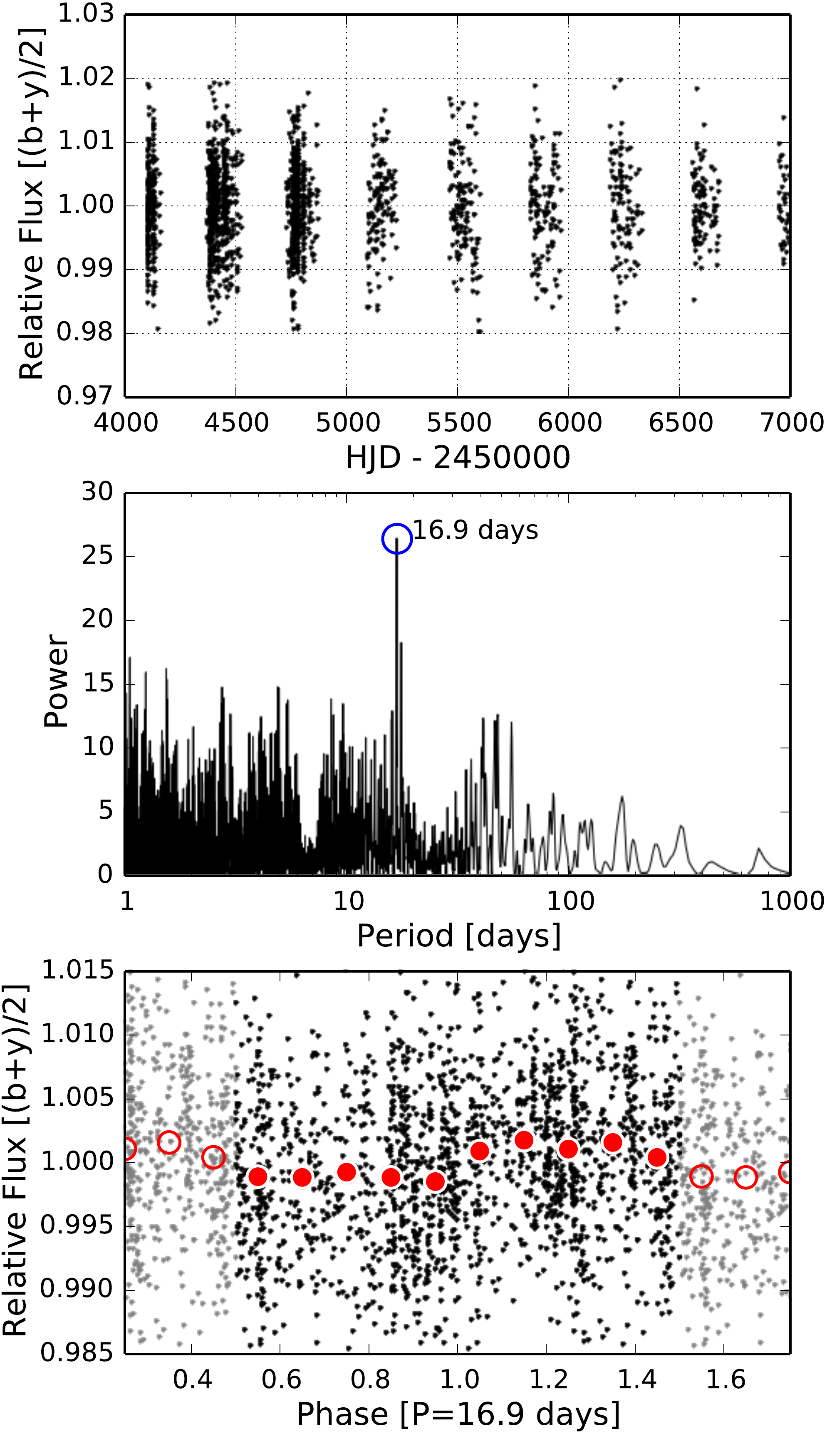}
         \end{center}
         \caption{
              Differential photometry of HD 7924 from APT.
              \emph{Top:} Relative flux time-series of HD 7924 in the combined Str\"{o}mgren $(b+y)/2$ passband. Seasonal offsets are removed by dividing by the mean within each season. The standard deviation of the photometric time series is 2.3 parts per thousand (ppt).
              \emph{Middle:} LS periodogram of the photometric time series.
              \emph{Bottom:} Differential photometry from APT phase-folded to the 16.9 day peak found in the LS periodogram and binned to widths of 0.1 phase units (red circles). The light grey points and open red circles show the same data wrapped by one period. The peak-to-peak amplitude of the 16.9 day photometric signal is 1.1 ppt.
         }
         \label{fig:photperi}
\end{figure}

The 1855 photometric observations are plotted in the top panel of Figure \ref{fig:photperi} and summarized in Table \ref{tab:photometry}, where we have removed yearly offsets by dividing the observations in each season by the seasonal mean. We find a significant periodic signal in the full photometric dataset with a period of 16.9 days and peak-to-peak amplitude of 1.9 ppt. This is similar to the 17.1 day period found in the RVs and most likely a signature of the stellar rotation. Spots may form on opposite hemispheres and cause photometric fluctuations at one half of the orbital period. When the data are broken up into the individual observing seasons we find that the strongest photometric periodicity happens during the 2012-2013 season and has a period of 41.5 days. This season corresponds to the time of maximal \shk activity index. Starspots during this time likely dominate the rotationally induced signal in the RV data.

\begin{figure*}
         \begin{center}
              \includegraphics[width=1.25\textwidth]{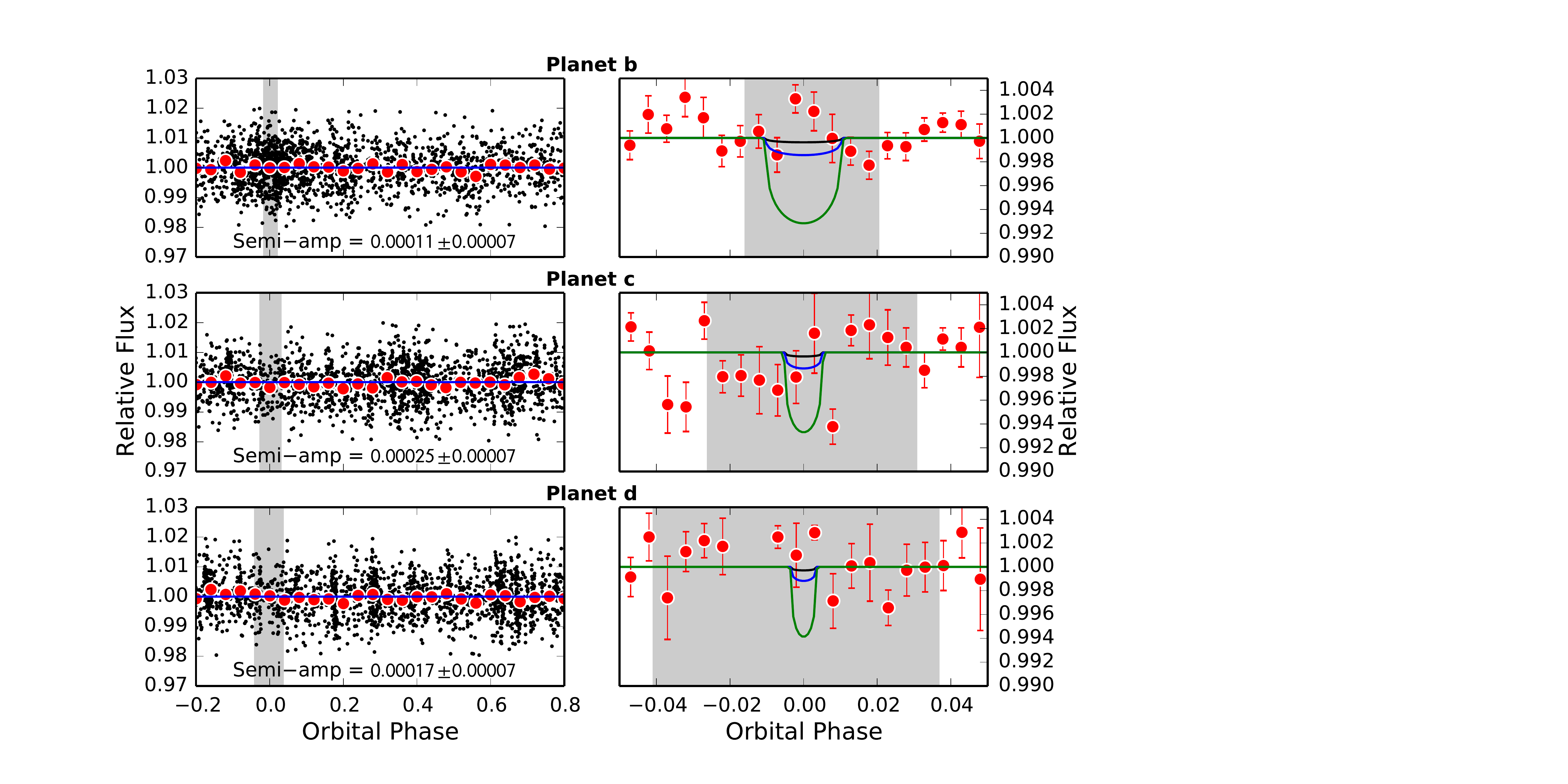}
         \end{center}
         \caption{
              Differential photometry from APT of HD 7924 phase-folded to the ephemerides of the three planets.
              \emph{Left:} Differential photometry phase-folded to the orbital period of planet b (top), c (middle), and d (bottom). Red circles are the photometry measurements grouped in bins of width 0.04 units of orbital phase. The grey shaded region spans the uncertainty in mid-transit time. The semi-amplitude of a sine wave least squares fit to the data is annotated at the bottom of each panel.
              \emph{Right:} Same as the left panels zoomed in around the phase of center transit. The individual measurements are omitted and the red circles correspond to binned photometry data in bins of width 0.005 units of orbital phase. The three curves are \citet{Mandel02} transit models for planets with the masses listed in Table \ref{tab:dparams} and densities of iron (black), water (blue), and hydrogen (green).
         }
         \label{fig:photphase}
\end{figure*}

No significant periodic variability is found in the photometry at the periods of the three planets. A least-squares sine wave fit to the data gives semi-amplitudes of 0.11 ppt, 0.25 ppt, and 0.17 ppt at the orbital periods of planets b, c, and d respectively. Figure \ref{fig:photphase} shows the photometric data phase-folded to the periods of each of the three planets. This gives further evidence that the 5.4, 15.3, and 24.5 day signals are planetary in nature and not caused by stellar activity that would be visible in the photometry at those periods.

With the APT data we are able to rule out transits deeper than $\approx$2 ppt for planet b and 5-8 ppt for planets c and d. We binned the data in bins of width 0.005 units of orbital phase around the time of center transit for each of the planets. For planet b, the lowest binned measurement within the uncertainty window of the transit time is $2.3\pm1.2$ below the mean suggesting that transits deeper than 3.5 ppt do not occur. With a mass of \bmass\ we can rule out non-grazing transits of a pure hydrogen planet, but transits of a denser planet would not be detectable in our data (see Figure \ref{fig:photphase}). Our constraints on possible transits of planets c and d are not as strong because their ephemerides have not yet been subject to intense campaigns to search for transits. However, the lowest binned measurements within the transit windows for planets c and d are $6.23\pm1.5$ ppt and $3.4\pm1.5$ ppt, respectively, and suggest that transits deeper than 7.7 ppt for planet c and 4.9 ppt for planet d are not present in the data. This rules out transits of a pure hydrogen planet d, but the lowest binned measurement for planet c is consistent with the depth of a transit caused by a pure hydrogen planet. However, if this were a real event we would expect the neighboring measurements to also be slightly lower than the mean since the transit duration is longer than the width of the bins but this is not the case. Both of the neighboring measurements are higher than the mean flux level. We conclude that transits of planets with compositions dominated by hydrogen for any of the three planets are unlikely, but more and/or higher precision observations are needed to exclude transits of rocky planets.

\begin{deluxetable*}{cccccc}
\tabletypesize{\small}
\tablewidth{0pt}
\tablecaption{SUMMARY OF APT PHOTOMETRIC OBSERVATIONS}
\tablehead{
\colhead{Observing} & \colhead{} & \colhead{Julian Date Range} &
\colhead{Sigma} & \colhead{$P_{\rm rot}$\tablenotemark{1}} & \colhead{Full Amplitude\tablenotemark{2}} \\
\colhead{Season} & \colhead{$N_{obs}$} & \colhead{(HJD $-$ 2,440,000)} &
\colhead{(mag)} & \colhead{(days)} & \colhead{(mag)}
}
\startdata
2006--2007 & 231 & 14100--14158 & 0.00212 & $17.1\pm0.2$ & $0.0021\pm0.0004$ \\
2007--2008 & 524 & 14370--14523 & 0.00215 & $17.1\pm0.1$ & $0.0011\pm0.0003$ \\
2008--2009 & 464 & 14728--14867 & 0.00208 & $45.7\pm0.5$ & $0.0013\pm0.0003$ \\
2009--2010 & 123 & 15092--15222 & 0.00212 &    \nodata   &      \nodata      \\
2010--2011 & 140 & 15459--15598 & 0.00239 & $18.1\pm0.1$ & $0.0024\pm0.0005$ \\
2011--2012 & 125 & 15823--15963 & 0.00222 & $16.8\pm0.1$ & $0.0021\pm0.0005$ \\
2012--2013 & 109 & 16185--16330 & 0.00235 & $41.5\pm0.8$ & $0.0027\pm0.0006$ \\
2013--2014 & 100 & 16555--16674 & 0.00168 & $25.1\pm0.3$ & $0.0022\pm0.0004$ \\
2006--2014 &1816 & 14100--16674 & 0.00197 & $16.8922\pm0.0014$ & $0.00098\pm0.00013$ \\
\enddata
\label{tab:photometry}
\tablenotetext{1}{Period of most significant peak in a periodogram analysis.}
\tablenotetext{2}{Amplitude of best-fit sine function with the period fixed to $P_{\rm rot}$.}
\end{deluxetable*}

\section{Spitzer Transit search}
\begin{figure}
         \begin{center}
              \includegraphics[width=0.5\textwidth]{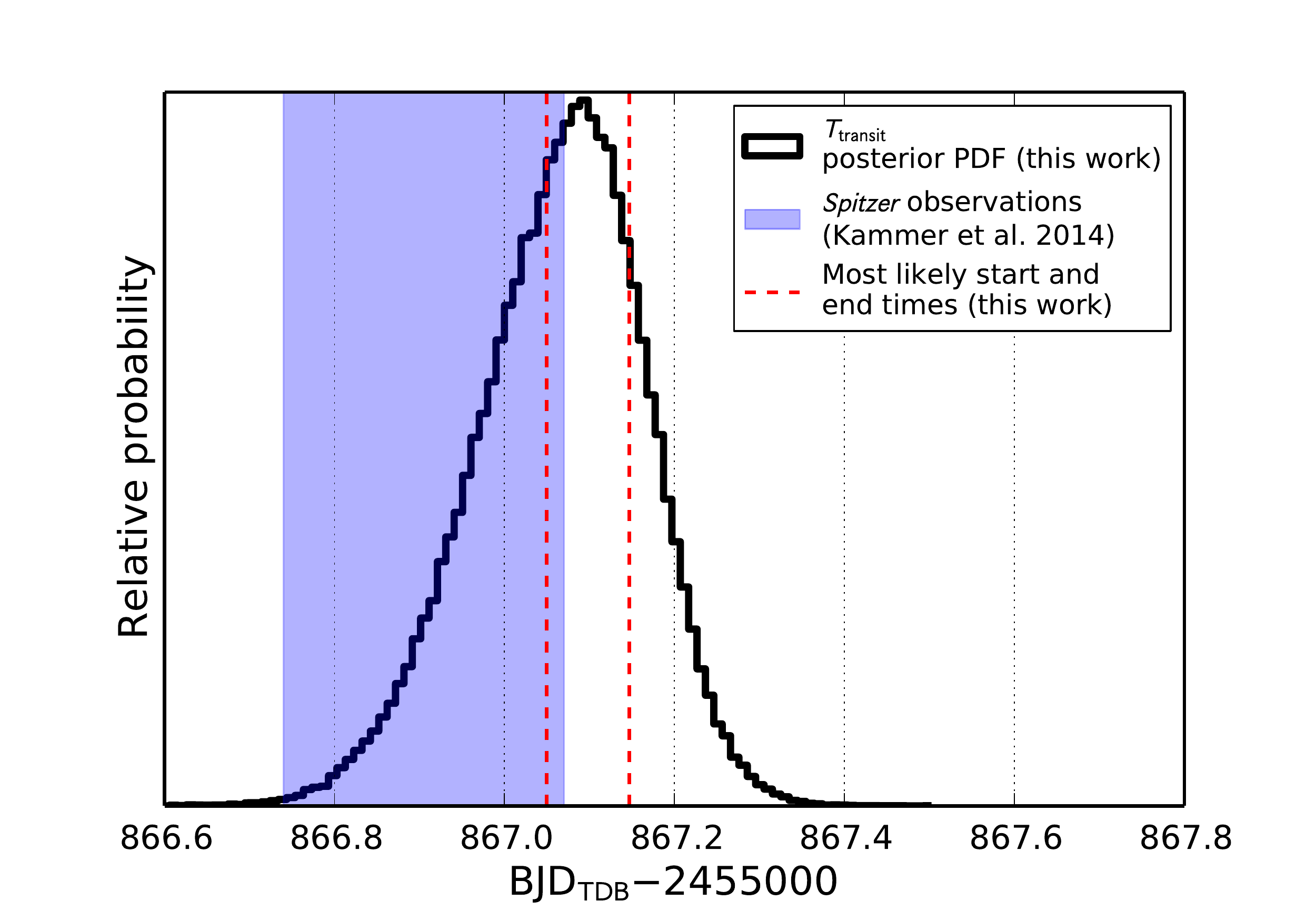}
         \end{center}
         \caption{
              Comparison of the transit search time window covered by the \emph{Spitzer} observations of \citet{Kammer14} to the transit times as calculated from the updated ephemeris in this work. The thick black line shows the posterior probability density function (PDF) of the mid-transit time for the transit that was targeted by \citet{Kammer14}. The blue shaded region shows the time window covered by their observations, and the red dashed lines show the most likely ingress and egress times.
         }
         \label{fig:tran}
\end{figure}

A photometric campaign to look for transits of HD 7924b using the \emph{Spitzer} Space Telescope showed no evidence of transiting planets larger than 1.16 \rearthe \citep[$2-\sigma$ confidence,][]{Kammer14}. However, they assumed a circular orbit which caused them to underestimate the uncertainty on the time of transit. We find an ephemeris that is inconsistent with that of \citet{Kammer14} by $\sim3-\sigma$, but nearly identical to that of H09.
With the new ephemeris listed in Table \ref{tab:params} we calculate a time of transit for the epoch during which the \emph{Spitzer} observations were collected of $2455867.07^{+0.09}_{-0.11}$. The \citet{Kammer14} observations would have covered 70\% of the predicted ingress or egress times assuming a perfectly edge-on viewing angle for the orbit. The a priori transit probability for planet b is 6.4\%. Assuming the \citet{Kammer14} observations are of sufficient precision to detect any ingress or egress that would have occurred within their observational window, the a posteriori transit probability is 2.0\%.
Figure \ref{fig:tran} compares the observational window of the \emph{Spitzer} observations to the predicted time of transit based on the ephemeris of this work.

\section{Architecture and Stability}
\subsection{Compact Multi-planet Systems}
\begin{figure}
         \begin{center}
              \includegraphics[width=0.45\textwidth]{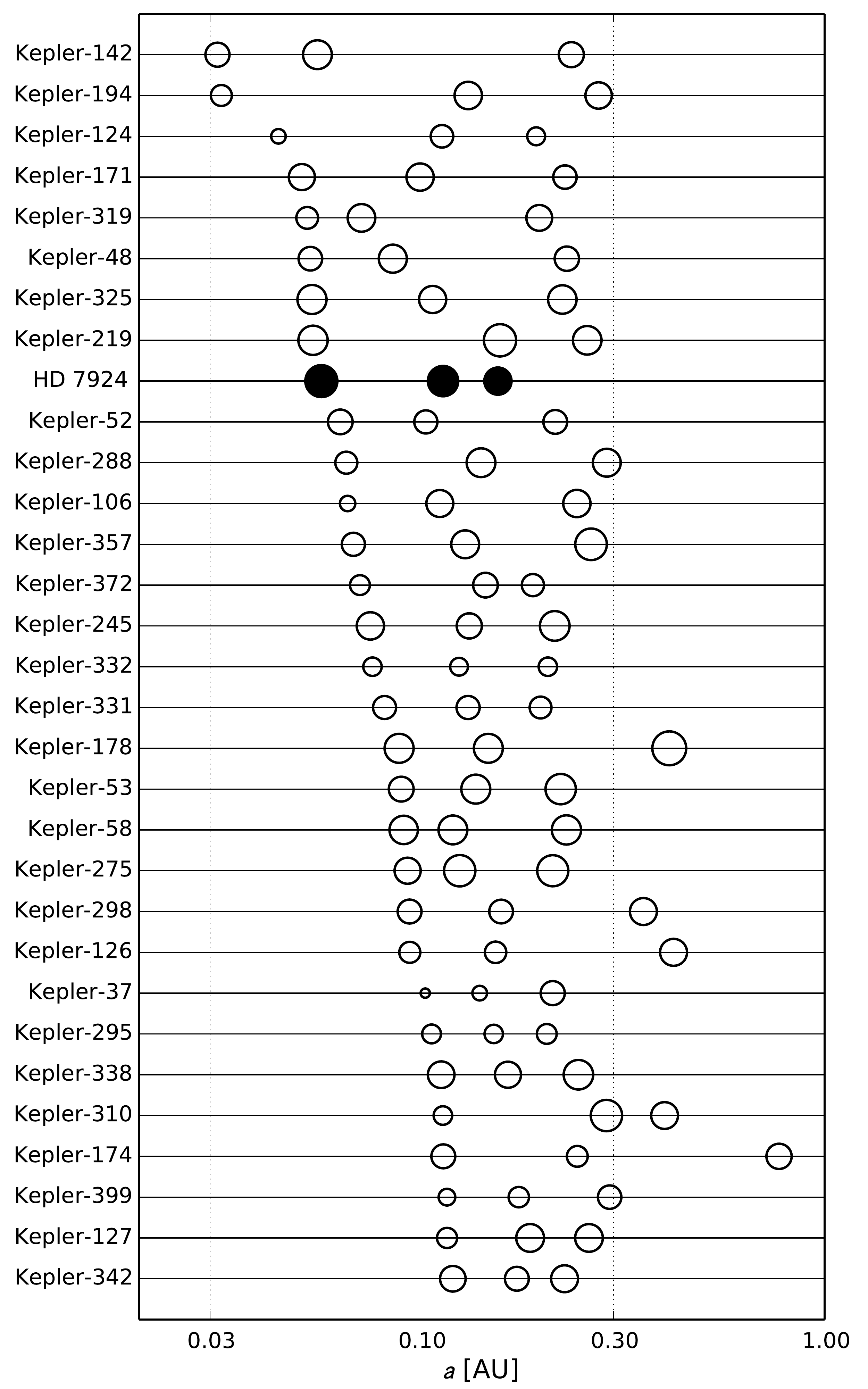}
         \end{center}
         \caption{
              The architecture of \emph{Kepler} multi-planet systems compared with the architecture of the HD 7924 system. We plot the \emph{Kepler} systems with exactly three currently known planets, a largest orbital period of 30 days or less, and a maximum planetary radius of 4 \rearthe. The HD 7924 planets are plotted as filled black circles. The systems are sorted by the semi-major axis of the shortest period planet in the system. The size of the circles is proportional to planet radius. The radii of the HD 7924 planets have been calculated as $\frac{R_{\rm p}}{R_{\oplus}}=0.371\left(\frac{M_{\rm p}}{M_\oplus}\right)^{1.08}$ \citep{Weiss14}.
         }
         \label{fig:multis}
\end{figure}

In order to compare the architecture of the HD 7924 system with other compact multi-planet systems we compiled a catalogue of similar multi-planet systems from \emph{Kepler}. We restricted the \emph{Kepler} multi-planet systems to those with exactly three currently known transiting planets in the system\footnote{As returned by a 2014 Nov 24 query of exoplanets.org}, a largest orbital period less than or equal to 30 days, and all three planets must have radii smaller than 4 \rearthe. This left a total of 31 systems from the sample of confirmed \emph{Kepler} systems \citep{Borucki11,Batalha13, Marcy14,Rowe14}. Their architectures are presented in Figure \ref{fig:multis}.

We notice nothing unusual about the architecture of HD 7924 when compared with the \emph{Kepler} systems. Most of these multi-planet systems contain three planets with masses between 5-10 \mearthe\ and semi-major axis between 0.05 and 0.3 AU. The systems of \emph{Kepler}-194, \emph{Kepler}-124, \emph{Kepler}-219, \emph{Kepler}-372, \emph{Kepler}-310, \emph{Kepler}-127, and \emph{Kepler}-342 all host one inner planet and two outer planets that are closer to each other than to the innermost planet as in the HD 7924 system. The \emph{Kepler}-372 system is particularly similar with one planet orbiting at 0.07 AU and a pair of outer planets orbiting at 0.14 and 0.19 AU. This demonstrates that the HD 7924 system is not abnormal in the context of other known compact multi-planet systems. While this is not a direct demonstration of dynamical stability, the fact that many other systems exist with very similar architectures gives strong empirical evidence that the HD 7924 system is a stable planetary configuration.

The HD 7924 planetary system is one of very few RV-detected systems hosting three super Earths. Only HD 40307 \citep{Mayor09a}, and HD 20794 \citep{Pepe11} are known to host three planets with masses all below 10 \mearthe. Both of these stars are unobservable from most northern hemisphere observatories. If we expand the mass limit to include systems with three or more planets with masses below 25 \mearth we find four additional systems that match these criteria, HD 69830 \citep{Lovis06}, Gl 581 \citep{Mayor09b}, HD 10180  \citep{Lovis11}, and 61 Virginis \citep{Vogt10}. These systems are difficult to detect given their small and complex RV signals and HD 7924 is the first discovery of such a system since 2011.

\subsection{Dynamical Stability}
In order to check that the HD 7924 system is dynamically stable for many orbital periods we ran a numerical integration of the three planet system using the \emph{MERCURY} code \citep{Chambers99}. We started the simulation using the median orbital elements presented in Table \ref{tab:params} and let it proceed $10^5$ years into the future. We assume that the system is perfectly coplanar with zero mutual inclination and we assume that the masses of the planets are equal to their minimum masses ($\sin{i} = 0$). No close passages between any of the three planets were found to occur ($\leq$1 Hill radius) during the entire simulation, suggesting that the system is stable in the current configuration.

Following the arguments of \citet{Fabrycky14} we calculated the sum of separations between the planets (b to c, plus c to d) in units of their Hill radii. \citet{Fabrycky14} found that the sum of the separations for the vast majority of \emph{Kepler} multi-planet systems is larger than 18 Hill radii. The sum of the separations between the planets of the HD 7924 system is 36 Hill radii, providing further evidence that this system is stable and not abnormally compact relative to the many \emph{Kepler} multi-planet systems.

\section{APF vs.\ Keck}
\label{sec:apfvkeck}

Since this is the first publication from our group utilizing APF data we perform a comparison between the two data sets to assess their relative performance. First, we use the automated planet detection algorithm described in \S \ref{sec:disc} on each data set independently. When the APF and Keck data are analyzed independently the 5.4, 15.3, and 24.5 day signals are the first to be identified in both cases (albeit with much lower significance in the APF data alone) making it extremely unlikely that these signals could be caused by instrumental systematic noise.

\begin{figure*}
         \begin{center}
              \includegraphics[width=0.8\textwidth]{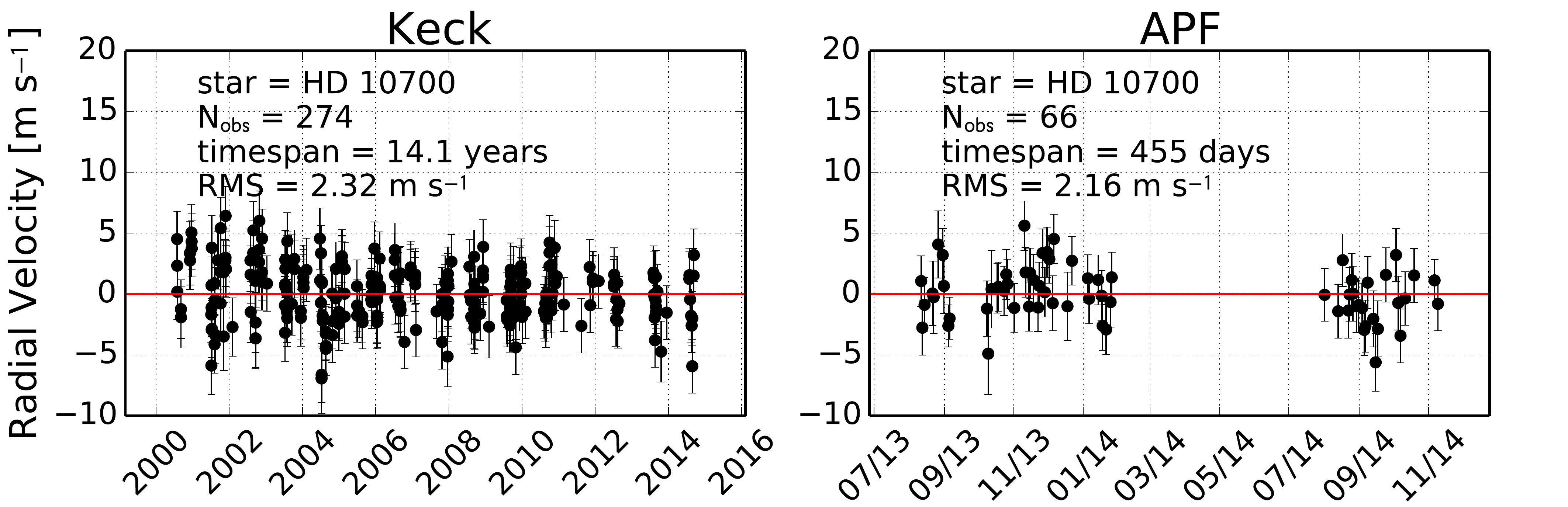}
              \includegraphics[width=0.8\textwidth]{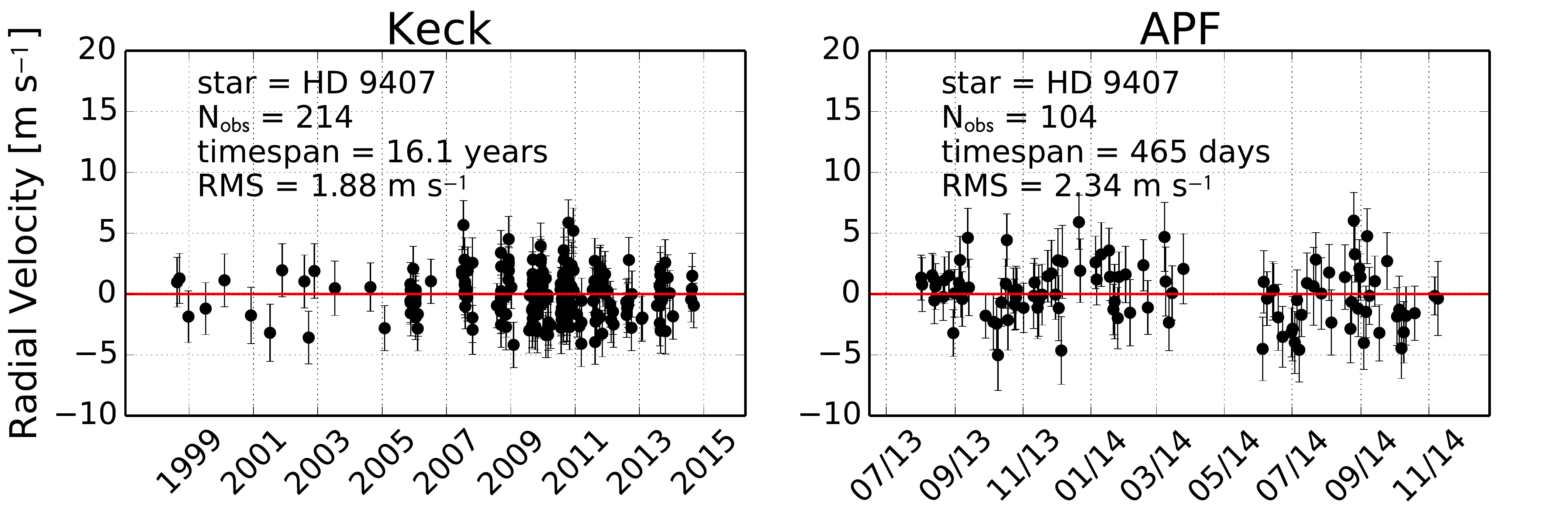}
         \end{center}
         \caption{
              A comparison between RV performance at Keck and APF for two well known RV standard stars. These stars are some of the most well-observed stars at both Keck and APF. In the first 1.5 years of APF operations we have already collected nearly half of the data that has been collected at Keck over the last $\sim$15 years. Stellar jitter has been added in quadrature with the binned velocities such that the $\chi^2$ of the velocities with respect to their median value is 1.0. 
         }
         \label{fig:apfvkeck}
\end{figure*}

The binned measurement uncertainties for the APF data are generally higher than those of Keck (before stellar jitter is accounted for). This is not surprising since the APF data is usually collected at lower signal to noise than the Keck data. However, since we are allocated many more nights on APF we can collect more measurements over a shorter amount of time. It took $\sim$13 years to collect \binkeck\ binned measurements with Keck, but only 1.5 years to collect 80 binned measurements with the APF. In addition, HD 7924's circumpolar track at Lick Observatory and the long nights during the winter months in Northern California both increase the observability of this target. The standard deviation of the residuals to the best fit model are similar in the two cases at 2.8 \ms\ for APF and 2.5 \ms\ for the post-upgrade Keck data. In order to compare the tradeoffs between higher cadence and higher precision we define a metric similar to the ``photometric noise rate" as used by \citet{Fulton11} to compare the relative performance between photometric data sets. We define a ``velocity noise rate" as
\begin{equation}
\label{eqn:vnr}
\textrm{VNR} = \frac{\sigma}{\sqrt{\Gamma}},
\end{equation}
where $\sigma$ is the RMS of the velocities and $\Gamma$ is the mean number of observations per year. We assume that no further signals are present in the data and/or they contribute negligibly to the RMS. The VNR gives the approximate $K$ in \ms\ that would be detectable with S/N=1 after one year of observing at typical cadence. Of course, we generally require S/N $\gg1$ in order to consider a signal a viable planet candidate, but the VNR still gives a good reference point for comparison. We list the VNR and other performance characteristics for Keck and APF data for HD 7924 and two well-known RV standard stars (HD 10700, and HD 9407) in Table \ref{tab:performance}. The velocities for the three stars are compared side-by-side in Figure \ref{fig:apfvkeck}. The VNR for APF data is 25--50\% smaller than that of Keck indicating that we will be sensitive to smaller planets once we have observed the stars for a comparable amount of time or, in other words, our APF data will be of comparable sensitivity to the existing Keck RVs in 1/2 to 3/4 of the length of time it took to reach that sensitivity at Keck.

We are actively tuning and perfecting our Doppler analysis pipeline for APF. Recently, we implemented an experimental technique to correct the velocities for any correlations of the RVs with environmental parameters (e.g. atmospheric pressure, CCD temperature). This technique reduces the RV RMS for Keck data slightly (5-10\%), but significantly reduces the RMS for APF velocities in most cases and up to a factor of two in some cases. The RV RMS for HD 9407 and HD 10700 reduce to 2.24 \ms and 1.67 \ms respectively when correlations with non-astrophysical variables are removed.

\begin{deluxetable*}{lccrrc}
\tabletypesize{\footnotesize}
\tablecaption{APF vs.\ Keck Radial Velocity Precision}
\tablehead{ 
    \colhead{Instrument}   & \colhead{RMS}  & \colhead{Median Uncertainty\tablenotemark{1}} & \colhead{$N_{\rm obs}$\tablenotemark{2}} &  \colhead{Mean Cadence} & \colhead{VNR\tablenotemark{3}} \\
    \colhead{}  & \colhead{(\ms)}            & \colhead{(\ms)}       & \colhead{} & \colhead{(days)} & \colhead{\ms y$^{-1}$}
}
\startdata
\sidehead{{\bf HD 7924}}
Pre-upgrade Keck/HIRES        & 1.20 & 2.64 &     7 & 110.4 & 0.66 \\
Post-upgrade Keck/HIRES      & 2.50 & 2.36 & \binkeck &   13.5 & 0.48 \\
APF/Levy                                 & 2.80 & 2.86 &   \binapf &     5.9 & 0.36 \\
\hline

\sidehead{{\bf HD 10700 ($\tau$ Ceti)}}
Pre-upgrade Keck/HIRES        & 2.87 & 2.84 &     84 &   17.5 & 0.63 \\
Post-upgrade Keck/HIRES      & 2.32 & 2.22 &   190 &   19.4 & 0.53 \\
APF/Levy                                 & 2.16 & 2.12 &     66 &     7.0 & 0.30 \\
\hline

\sidehead{{\bf HD 9407}}
Pre-upgrade Keck/HIRES        & 1.89 & 1.94 &     12 & 162.8 & 1.26 \\
Post-upgrade Keck/HIRES      & 1.89 & 1.86 &   202 &   18.3 & 0.52 \\
APF/Levy                                 & 2.34 & 2.30 &   104 &     4.5 & 0.26 \\
\hline
\enddata
\tablenotetext{1}{Stellar jitter has been added in quadrature with the binned velocities such that the $\chi^2$ of the velocities with respect to their median value is 1.0.}
\tablenotetext{2}{Binned in units of 0.5 days.}
\tablenotetext{3}{See Equation \ref{eqn:vnr}.}
\label{tab:performance}
\end{deluxetable*}

\section{Discussion \& Summary}
\label{sec:discussion}
We present the discovery of two additional super-Earth mass planets orbiting the bright K0.5 dwarf HD 7924. These planets join a previously-known planet in a system with at least three super-Earth mass planets. The two new planets have minimum masses of $M_{c}\sin{i_c} = 7.8$ \mearthe\ and $M_{d}\sin{i_d} = 6.3$ \mearthe, and orbit HD 7924 with semi-major axis of $a_{c} = 0.113$ AU and $a_{d} = 0.155$ AU. Both planets receive far too much radiation from HD 7924 to be within the habitable zone as defined by \citet{Kopparapu13} with incident stellar irradiation values 114, 28, and 15 times that received from the Sun by Earth for planets b, c, and d respectively. Assuming that these planets have bond albebos similar to the mean total albedos of super-Earths \citep[$A_{t}=0.32$,][]{Demory14} their equilibrium temperatures are 826, 584, and 499 K.

The stellar magnetic activity cycle is clearly visible in our long-baseline RV time series and we observe nearly two complete cycles. We simultaneously model the RV shift due to the magnetic cycle and the three planets in order to extract accurate planetary parameters. A tentative RV signal from rotationally-modulated starspots is also found and we perform a rigorous analysis to determine that the planetary signals are distinct from the stellar activity signals.

With the largely expanded data set we are able to refine the ephemeris of planet b and show that additional transit monitoring is needed. Since HD 7924 is near the ecliptic pole, it will be near the continuous viewing zone for the James Webb Space Telescope \citep{Gardner06} and a transiting planet would make an excellent candidate for space-based transmission spectroscopy. The Transiting Exoplanet Survey Telescope \citep{Ricker14} will observe HD 7924 once launched and will be able to conclusively determine if any of the three planets are transiting.

This system is a good example of a compact system of short-period planets for which high-cadence observations are incredibly valuable to determine the coherence of signals and detect the true physical periods as opposed to aliases. Since we know that short-period super-Earth size planets are common \citep{Howard10}, we expect our continued nearly nightly observations with the APF to uncover many more systems like HD 7924 from which we will build a comprehensive census of the small planets in our local solar neighborhood.

\acknowledgments{
We are very grateful for the donations of the Levy family that helped facilitate the construction of the Levy spectrograph on APF. Without their support the APF would not be contributing to the discovery of planets like these. In their honor we informally name the HD 7924 system ``The Levy Planetary System".
We thank the many observers who contributed to the measurements reported here.
We thank Kyle Lanclos, Matt Radovan, Will Deich and the rest of the UCO Lick staff for their invaluable help shepherding, planning,
and executing observations, in addition to writing the low-level software that made the automated APF observations possible.
We thank Debra Fischer, Jason Wright, and John Johnson for their many nights of observing that contributed to the Keck data presented in this work.
We gratefully acknowledge the efforts and dedication of the Keck Observatory staff, 
especially Scott Dahm, Greg Doppman, Hien Tran, and Grant Hill for support of HIRES 
and Greg Wirth for support of remote observing.  
We are grateful to the time assignment committees of the University of Hawai`i, the University of California, and NASA 
for their generous allocations of observing time.  
Without their long-term commitment to RV monitoring, these planets would likely remain unknown.  
We acknowledge R.\ Paul Butler and S.\,S.\ Vogt for many years
of contributing to the data presented here.
A.\,W.\,H.\ acknowledges NASA grant NNX12AJ23G.
L.\,M.\,W.\ gratefully acknowledges support from Ken and Gloria Levy.
This material is based upon work supported by the National Science Foundation Graduate Research Fellowship under Grant No. 2014184874. Any opinion, findings, and conclusions or recommendations expressed in this material are those of the authors and do not necessarily reflect the views of the National Science Foundation.
G.\,W.\,H.\ acknowledges support from NASA, NSF, Tennessee State University, and
the State of Tennessee through its Centers of Excellence program.
This research made use of the Exoplanet Orbit Database
and the Exoplanet Data Explorer at exoplanets.org.
This work made use of the SIMBAD database (operated at CDS, Strasbourg, France), 
NASA's Astrophysics Data System Bibliographic Services, 
and the NASA Star and Exoplanet Database (NStED).
Finally, the authors wish to extend special thanks to those of Hawai`ian ancestry 
on whose sacred mountain of Maunakea we are privileged to be guests.  
Without their generous hospitality, the Keck observations presented herein
would not have been possible.
}

\newpage
\bibliographystyle{apj}
\bibliography{hd7924_3pl}

\enddocument